\documentclass[final,journal,letterpaper,twocolumn]{IEEEtran}
\usepackage[utf8]{inputenc}
\usepackage{amsmath}
\usepackage{graphicx}
\usepackage{float}
\usepackage{authblk}
\usepackage{subfig}
\usepackage{siunitx}
\usepackage{xcolor}
\usepackage{makecell}
\usepackage{multirow}
\usepackage[normalem]{ulem}
\usepackage{booktabs}

\usepackage[colorlinks=true,linkcolor=black,anchorcolor=black,citecolor=black,menucolor=black,runcolor=black,urlcolor=black,bookmarks=true]{hyperref}

\newcommand{\YS}[1]{\textcolor{black}{#1}}

\newcommand{\XZ}[1]{\textcolor{black}{#1}}
\newcommand{\YW}[1]{\textcolor{black}{#1}}

\newcommand{\SM}[1]{\textcolor{black}{#1}}

\newcommand{\SMrev}[1]{\textcolor{black}{#1}}

\newcommand{\MAN}[1]{\textcolor{black}{#1}}
\newcommand{\MA}[1]{\textcolor{black}{#1}}

\begin{document}


\title{Deep Learning Meets SAR}

\author{Xiao Xiang Zhu,~\IEEEmembership{Fellow,~IEEE},
        Sina Montazeri,~
        Mohsin Ali,~
        Yuansheng Hua,~\IEEEmembership{Member,~IEEE},
        Yuanyuan Wang,~\IEEEmembership{Member,~IEEE},
        Lichao Mou,~\IEEEmembership{Member,~IEEE},
        Yilei Shi,~\IEEEmembership{Member,~IEEE},
        Feng Xu,~\IEEEmembership{Senior Member,~IEEE},
        Richard Bamler,~\IEEEmembership{Fellow,~IEEE}

\thanks{The work of X. Zhu is jointly supported by the European Research Council (ERC) under the European Union's Horizon 2020 research and innovation programme (grant agreement No. [ERC-2016-StG-714087], Acronym: \textit{So2Sat}), by the Helmholtz Association
through the Framework of Helmholtz AI - Local Unit ``Munich Unit @Aeronautics, Space and Transport (MASTr)'' and Helmholtz Excellent Professorship ``Data Science in Earth Observation - Big Data Fusion for Urban Research'' and by the German Federal Ministry of Education and Research (BMBF) in the framework of the international future AI lab "AI4EO" (Grant number: 01DD20001).
(\emph{Corresponding author: Xiao Xiang Zhu}).}
\thanks{X. Zhu, M.~Ali, Y.~Hua, and L.~Mou are with the Remote Sensing Technology Institute (IMF), German Aerospace Center (DLR), Germany and with Data Science in Earth Observation (SiPEO, former: Signal Processing in Earth Observation), Technical University of Munich (TUM), Germany (e-mails: xiaoxiang.zhu@dlr.de).}
\thanks{S.~Montazeri and Y.~Wang are with DLR-IMF, Wessling, Germany.}
\thanks{Y.~Shi is with the Chair of Remote Sensing Technology (LMF), TUM, 80333 Munich, Germany.}
\thanks{F.~Xu is with the Key Laboratory for Information Science of Electromagnetic Waves (MoE), Fudan Univeristy, Shanghai, China.}
\thanks{R.~Bamler is  with DLR-IMF, 82234 Wessling, Germany and TUM-LMF, 80333 Munich, Germany.}}

\markboth{Accepted by IEEE GEOSCIENCE AND REMOTE SENSING MAGAZINE, 2021}%
{xxx \MakeLowercase{\textit{et al.}}: }

\maketitle

\begin{abstract}
\textcolor{blue}{This is the pre-acceptance version, to read the final version please go to IEEE Geoscience and Remote Sensing Magazine on IEEE XPlore.}

Deep learning in remote sensing has become an international hype, but it is mostly limited to the evaluation of optical data. Although deep learning has been introduced in Synthetic Aperture Radar (SAR) data processing, despite successful first attempts, its huge potential remains locked. In this paper, we provide an introduction to the most relevant deep learning models and concepts, point out possible pitfalls by analyzing special characteristics of SAR data, review the state-of-the-art of deep learning applied to SAR in depth,  summarize available benchmarks, and recommend some important future research directions. With this effort, we hope to stimulate more research in this interesting yet under-exploited research field \textcolor{black}{and to pave the way for use of deep learning in big SAR data processing workflows.}
\end{abstract}


\section{Motivation}

In recent years, deep learning \cite{lecun_deep_2015} has been developed at a dramatic pace, achieving great success in many fields. Unlike conventional algorithms, deep learning-based methods commonly employ hierarchical architectures, such as deep neural networks, to extract feature representations of raw data for numerous tasks. For instance, convolutional neural networks (CNNs) are capable of learning low- and high-level features from raw images with stacks of convolutional and pooling layers, and then applying the extracted features to various computer vision tasks, such as large-scale image recognition  \cite{simonyan2014very}, object detection \cite{zhao_object_2019}, and semantic segmentation \cite{guo_review_2018}.

Inspired by numerous successful applications in the computer vision community, the use of deep learning in remote sensing is now obtaining wide attention \cite{zhu2017deep}. As first attempts in Synthetic Aperture Radar (SAR), deep learning-based methods have been adopted for a variety of tasks, including terrain surface classification \cite{parikh_classification_2020}, object detection \cite{chen2014sar},  parameter inversion \cite{wang2015ice}, despeckling \cite{wang2017sar}, specific applications in Interferometric SAR (InSAR) \cite{anantrasirichai_application_2018}, and SAR-optical data fusion \cite{hughes2018identifying}. 

For terrain surface classification from SAR and Polarimetric SAR (PolSAR) images, effective feature extraction is essential. These features are extracted based on expert domain knowledge and are usually applicable to a small number of cases and data sets. Deep learning feature extraction has however proved to overcome, to some degrees, both of the aforementioned issues \cite{parikh_classification_2020}. For SAR target detection, conventional approaches mainly rely on template matching, where specific templates are created manually \cite{ikeuchi_invariant_1996} to classify different categories, or through the use of traditional machine learning approaches, such as Support Vector Machines (SVMs) \cite{zhao2001support, bryant1999svm}; in contrast, modern deep learning algorithms aim at applying deep CNNs to extract discriminative features automatically for target recognition \cite{chen2014sar}. For parameter inversion, deep learning models are employed to learn the latent mapping function from SAR images to estimated parameters, e.g., sea ice concentration \cite{wang2015ice}. Regarding despeckling, conventional methods often rely on artificial filters and may suffer from mis-eliminating sharp features when denoising. Furthermore, the development of joint analysis of SAR and optical images has been motivated by the capacities of extracting features from both types of images. For applications in InSAR, only a few studies have been carried out such as the work described in \cite{anantrasirichai_application_2018}. However, these algorithms neglect the special characteristics of phase and simply use an out-of-the-box deep learning-based model.
\par
Despite the first successes, and unlike the evaluation of optical data, the huge potential of deep learning in SAR and InSAR remains locked. For  example,  to  the  best  knowledge  of  the authors, there is no single example of deep learning in SAR that has  been  developed  up  to  operational  processing  of  big data  or integrated  into  the  production  chain  of  any  satellite  mission. This paper aims at  stimulating more research in this interesting yet under-exploited research field.
\par
In the remainder of this paper, Section \ref{sec:intro_dl} first introduces the most commonly used deep learning models in remote sensing. Section \ref{sec:pitfalls} describes the specific characteristics of SAR data that have to be taken into account to exploit the full potential of SAR combined with deep learning. Section \ref{sec:app} details recent advances in the utilization of deep learning on different SAR applications, which were outlined earlier in the section. Section \ref{sec:data} reviews the existing benchmark data sets for different applications of SAR and their limitations. Finally, Section \ref{sec:con} concludes current research, and gives an overview of promising future directions.

\vspace{1em}

\section{Introduction to Relevant Deep Learning Models and concepts}
\label{sec:intro_dl}

\begin{figure*}[ht]
\centering
\includegraphics[width=1\textwidth]{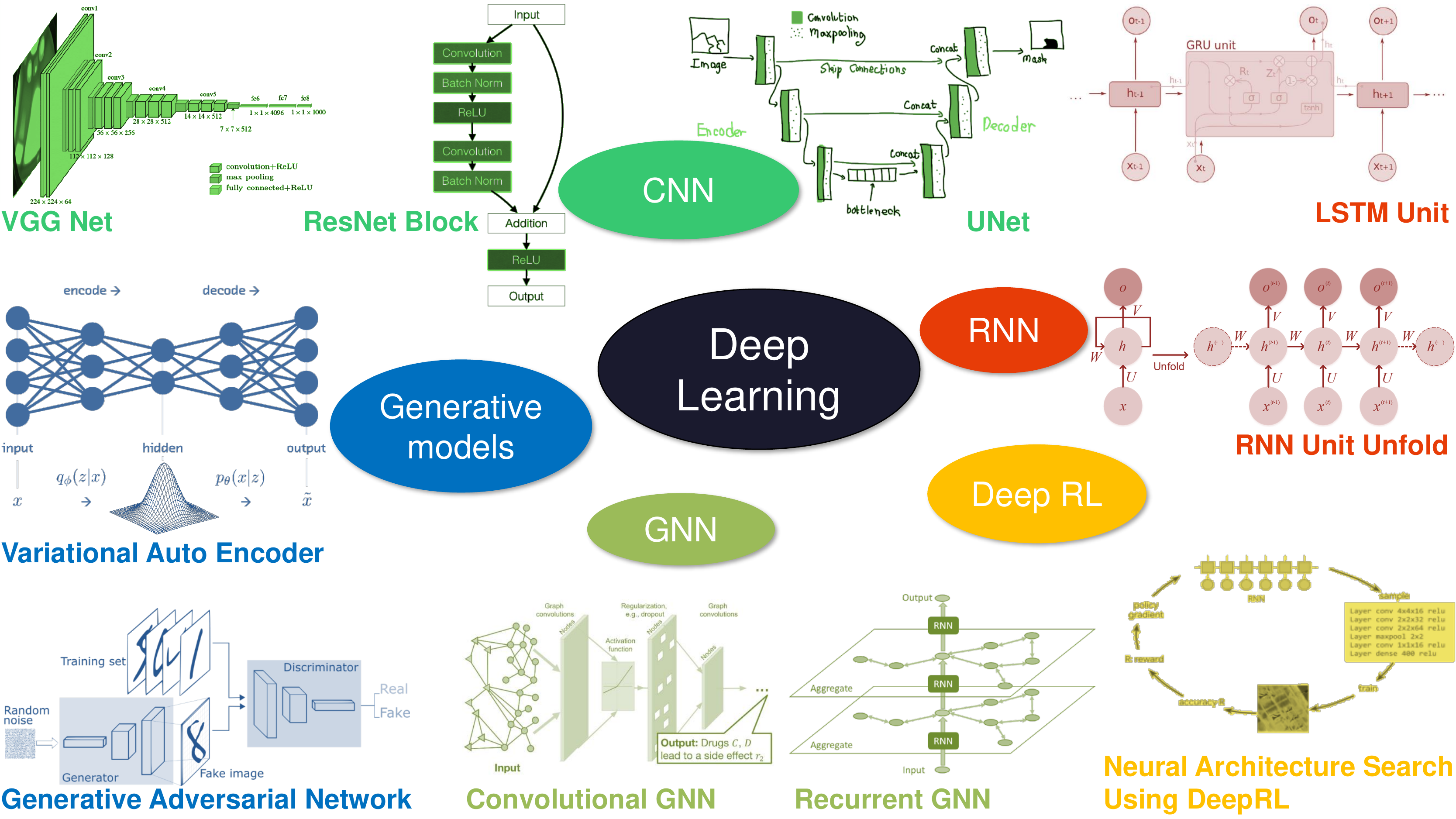}
\caption{A Selection of relevant deep learning models. Sources of the images: VGG \cite{ferguson_automatic_2017}, ResNet \cite{online_resnet}, U-Net \cite{online_unet}, LSTM \cite{online_lstm}, RNN \cite{feng_audio_2017}, VAE \cite{online_vae}, GAN \cite{online_gan}, CGNN \cite{Zitnik2018}, RGNN \cite{huang_residual_2019}, and DeepRL \cite{zoph_neural_2017}.}
\label{fig:graphic_abstract}
\end{figure*}

\par
In this section, we briefly review relevant deep learning algorithms originally proposed for visual data processing that are widely used for the state-of-the-art research of deep learning in SAR. In addition, we mention the latest developments of deep learning, which are not yet widely applied to SAR but may help create next generation of its algorithms. Fig.~\ref{fig:graphic_abstract} gives an overview of the deep learning models we discuss in this section.

Before discussing deep learning algorithms, we would like to stress that the importance of high-quality benchmark datasets in deep learning research cannot be overstated. Especially in supervised learning, the knowledge that can be learned by the model is bounded by \SM{the} information present in the training dataset. For example, the MNIST \cite{lecun2010mnist} dataset played \SM{a} key role in \SM{Yann LeCun's} seminal paper about convolutional neural networks and gradient-based learning \cite{lecun1998gradient}. Similarly, there would be no \SM{AlexNet} \cite{krizhevsky2012imagenet}, the network that kick-started the current deep learning renaissance, without \SM{the} ImageNet \cite{deng2009imagenet} dataset, which contains over 14 million images and 22,000 classes. ImageNet has been such \SM{an} important part of deep learning research that, even after over 10 years of being published, it is still used as a standard benchmark to evaluate the performance of CNNs for image classification.

\subsection{Deep Learning Models}
The main principle of deep learning models is to encode input data into effective feature representations for target tasks. To examplify how a deep learning framework works, we take autoencoder as an example: it first maps an input data to a latent representation via a trainable nonlinear mapping and then reconstructs inputs through reverse mapping. The reconstruction error is usually defined as the Euclidian distance between inputs and reconstructed inputs. Parameters of autoencoders are optimized by gradient descent based optimizers, like stochastic gradient descent (SGD), RMSProp \cite{Tieleman2012} and ADAM \cite{kingma2014adam}, during the backpropagation step.


\subsubsection{Convolutional Neural Networks (CNN)}
With the success of AlexNet in the ImageNet Large Scale Visual Recognition  Challenge (ILSVRC-2012), where it scored a top-5 test error of 15.3\% compared to 26.2\% of \SM{the} second best, CNNs have attracted worldwide attention and are now used for \SM{many} image understanding tasks, such as image classification, object detection, and semantic segmentation. AlexNet consists of five convolutional layers, three max-pooling layers, and three fully-connected layers. One of the key innovations of the AlexNet was the use of GPUs, which made it possible to train such large networks with huge datasets without using supercomputers. In just two years, VGGNet \cite{simonyan2014very} overtook AlexNet in performance by achieving a 6.8\% top-5 test error in ILSVRC-2014; the main difference was that it only used 3x3-sized convolutional kernels, which enabled it to have more number of channels and in turn capture more diverse features.

ResNet \cite{he2016deep}, U-Net \cite{ronneberger2015u}, and \SM{DenseNet} \cite{huang2017densely} were the next major CNN architectures. The main feature of all these architectures was the idea of connecting, not only neighboring layers but any two layers in the network, by using skip connections. This helped reduce loss of information across networks, mitigated the problem of vanishing gradients and allowed the design of deeper networks. U-Net is one of the most commonly used image segmentation networks. It has autoencoder based architecture where it uses skip connections to concatenate features from the first layer to last, second to second last, and so on: this way it can get fine-grained information from initial layers to the end layers. U-Net was initially proposed for medical image segmentation, where data labeling is a big problem. The authors used heavy data augmentation techniques on input data, making it possible to learn from only a few hundred annotated samples. In ResNet skip connections were used within individual blocks and not across the whole network. Since its initial proposal, it has seen many architectural tweaks, and even after 4-5 years its variants are always among \SM{the} top scorers on ImageNet. In DenseNet all the layers were attached to all preceding layers, reducing the size of the network, albeit at the cost of memory usage. 
For a more detailed explanations of different CNN models, interested readers are referred to \cite{hoeser_object_2020}. \MAN{These CNN models have also proved their worth in SAR processing tasks e.g. see \cite{mazza_tandem-x_2019, lattari_deep_2019, morgan2015deep}. For more examples and details of CNNs in SAR we refer our readers to Section \ref{sec:app}}.

\subsubsection{Recurrent Neural Networks (RNN)}

Besides CNNs, RNNs \cite{pearlmutter1989learning} are another major class of deep networks. Their main building blocks are recurrent units, which take the current input and output of the previous state as input. They provide state-of-the-art results for processing data of variable lengths like text and time series data. Their weights can be replaced with convolutional kernels for visual processing tasks such as image captioning and predicting future frames/points in visual time-series data. Long short term memory (LSTM) \cite{hochreiter1997long} is one of the most popular architectures of RNN: its cells can store values from any past instances while not being severely affected by the problem of gradient diminishing. \MAN{Just like in any other time series data from deep learning toolkit the RNNs are natural choice to process SAR time series data, e.g. see \cite{ndikumana_deep_2018}}. 

\subsubsection{GANs}
Proposed by Ian Goodfellow et al. \cite{goodfellow2014generative}, GANs are among the most popular and exciting inventions in the field of deep learning. Based on game-theoretic principles, they consist of two networks called a generator and a discriminator. The generator's objective is to learn a latent space, through which it can generate samples from the same distribution as the training data, while the discriminator tries to learn to distinguish if a sample is from the generator or  training data. This very simple mechanism is responsible for most cutting-edge algorithms of various applications, e.g., generating artificial photo-realistic images/videos, super-resolution, and text to image synthesis. \MAN{For example in the SAR domain GANs have already been successfully used in cloud removal applications \cite{grohnfeldi2018conditional, Ebel20}. The reader is referred to Section \ref{sec:app} for more examples.}

\subsection{Supervised, Unsupervised and Reinforcement Learning}
\subsubsection{Supervised Learning}
Most of popular deep learning models fall under the category of supervised deep learning, i.e. they need labelled datasets to learn the objective functions. One of big challenges of supervised learning is generalization, i.e. how well a trained model performs on test data. Therefore it is vital that training data truly represents the true distribution of data so it can handle all the unseen data. If the model fits well on training data and fails on test data then it is called overfitting, in deep learning literature there are several techniques that can be used to avoid it, e.g. Dropout\cite{srivastava2014dropout}. 

\subsubsection{Unsupervised Learning}
Unsupervised learning refers to the class of algorithms where the training data do not contain labels. \SM{For instance,} in classical data analysis, principal component analysis (PCA) \cite{pearson1901liii} can be used to reduce \SM{the} data dimension \SM{followed by a clustering algorithm to group similar data points}. In deep learning generative models like autoencoders and variational autoencoders (VAEs) \cite{kingma2013auto} and Generative Adversarial Networks (GANs) \cite{goodfellow2014generative} are some of popular techniques that can be used for unsupervised learning. Their primary goal is to generate output data from the same distribution as input data. Autoencoders consists of an encoder part which finds compressed latent representation of input and a decoder part which decodes that representation back to the original input. VAEs take autoencoders to the next level by learning the whole distribution instead of just a single representation at the end of the encoder part, which in turn can be used by the decoder to generate the whole distribution of outputs. The trick to learning this distribution is to also learn variance along with mean of latent representation at the encoder-decoder meeting point and add a KL-divergence-based loss term to the standard reconstruction loss function of the autoencoders.

\subsubsection{Deep Reinforcement Learning (DeepRL)}
Reinforcement Learning (RL) tries to mimic the human learning behavior, i.e., taking actions and then adjusting them for the future according to feedback from the environment. For example, young children learn to repeat or not repeat their actions based on the reaction of their parents. The RL model consists of an environment with states, actions to transition between those states, and a reward system for ending up in different states. The objective of the algorithm is to learn the best actions for given states using a feedback reward system. In a classical RL algorithms function, approximators are used to calculate the probability of different actions in different states. DeepRL uses different types of neural networks to create these functions \cite{mnih2015human}\cite{mao2016resource}. Recently DeepRL received particular attention and popularity due to the success of Google Deep Mind's AlphaGo \cite{silver2016mastering}, which defeated the Go board game world champion. This task was considered impossible by computers just until a few years ago.

\subsection{Relevant Deep Learning Concepts}
\subsubsection{Automatic Machine Learning (AutoML)}
Deep networks have many hyperparameters to choose from, for example, number of layers, kernel sizes, type of optimizer, skip connections, \MA{and the like}. There are billions of possible combinations of these parameters and given high computational cost, time, and energy costs it is hard to find the best performing network even from among a few hundred candidates. In the case of deep learning, the objective of AutoML is mainly to find the most efficient and high performing deep network for a given dataset and task. The first major attempt in this field was by Zoph et al. \cite{zoph2016neural}, who used DeepRL to find the optimum CNN for image classification. In their system an RNN creates CNN architectures and, based on their classification results, proposes changes to them. This process continues to loop until the optimum architecture is found. This algorithm was able to find competing networks compared to the state-of-the-art but took over 800 GPUs, which was unrealistic for practical application. Recently, there have been many new developments in the AutoML field, which have made it possible to perform such tasks in more intelligent and efficient ways. More details about the field of network architectural search can be found in \cite{elsken2018neural}. \MAN{Furthermore AutoML have also already been successfully applied to SAR for PolSAR classification \cite{dong_automatic_2020}. The method shows great potential for segmentation and classification tasks in particular.}

\subsubsection{Geometric Deep Learning -- Graph Neural Networks (GNNs)}
Except for well-structured image data, there is a large amount of unstructured data, e.g., knowledge graphs and social networks, in real life that cannot be directly processed by a deep CNN. Usually, these data are represented in the form of graphs, where each node represents an entity and edges delineate their mutual relations. To learn from unstructured data, geometric deep learning has been attracting an increasing attention, and a most-commonly used architecture is GNN, which is also proven successful in dealing with structured data. Specifically, Using the terminology of graphs, nodes of a graph can be regarded as feature descriptions of entities, and their edges are established by measuring their relations or distances and encoded in an adjacency matrix. Once a graph is constructed, messages can be propagated among each node by simply performing matrix multiplication. Followingly, \cite{kipf2016semi} proposed Graph Convolutional Networks (GCNs) characterized by utilizing graph convolutions, and [45] fasten the process. Moreover recurrent units in RGNNs (Recurrent Graph Neural Network) \cite{huang1904residual} \cite{shi2020building} have also been proven to obtain achievements in learning from graphs. \MAN{The usefulness of GNNs in SAR is still to be properly explored as \cite{ma2019attention} is one of the only attempts in trying to do so.}

\section{Possible Pitfalls}
\label{sec:pitfalls}
To develop tailored deep learning architectures and prepare suitable training datasets for SAR or InSAR tasks, it is important to understand that SAR data is different from optical remote sensing data, not to mention images downloaded from the internet. In this section, we discuss the special characteristics (and possible pitfalls) encountered while applying deep learning to SAR.

What makes SAR data and SAR data processing by neural networks unique? SAR data are substantially different from optical imagery in many respects. These are a few points to be considered when transferring CNN experience and expertise from optical to SAR data:
\begin{itemize}
    \item \textbf{Dynamic Range.} Depending on their spatial resolution, the dynamic range of SAR images can be up to 90 dB (TerraSAR-X high resolution spotlight data with a resolution of about 1 m). Moreover, the distribution is extremely asymmetric, with the majority of pixels in the low amplitude range (distributed scatterers) and a long tail representing bright discrete scatterers, in particular in urban areas. Standard CNNs are not able to handle such dynamic ranges and, hence, most approaches feature dynamic compression as a preprocessing step. In \cite{tang2018sar}, the authors first take only amplitude values from 0 to 255 and then subtract mean values of each image.  In \cite{hughes2018identifying, mou2017cnn}, normalization is performed as a pre-processing step, which compresses the dynamic range significantly. 
    \vspace{1em}
    \item \textbf{Signal Statistics.} In order to retrieve features from SAR (amplitude or intensity) images the speckle statistics must be considered. Speckle is a multiplicative, rather than an additive, phenomenon. This has consequences: While the optimum estimator of radar brightness of a homogeneous image patch under speckle is a simple moving averaging operation (i.e., a convolution, like in the additive noise case), other optimum detectors of edges and low-level features under additive Gaussian noise may no longer be optimum in the case of SAR. A popular example is Touzi’s CFAR edge detector \cite{touzi1988statistical} for SAR images, which uses the ratio of two spatial averages over adjacent windows. This operation cannot be emulated by the first layer of a standard CNN.
    \par
    \vspace{0.5em}
    Some studies use a logarithmic mapping of the SAR images prior to feeding them into a CNN \cite{chierchia2017sar, wang2017sar}. This turns speckle into an additive random variable and \textemdash as a side effect \textemdash reduces dynamic range. But still, a single convolutional layer can only emulate approximations to optimum SAR feature estimators. It could be valuable to supplement the original log-SAR image by a few lowpass filtered and logarithmized versions as input to the CNN. Another approach is to apply some sophisticated speckle reduction filter before entering the CNN, e.g., non-local averaging \cite{shi2015optimized, zhu2014improving, denis_patches_2019}. 
    \vspace{1em}
    \item \textbf{Imaging Geometry.} The SAR image coordinates range and azimuth are not arbitrary coordinates like \textit{East} and \textit{North} or $x$ and $y$, but rather reflect the peculiarities of the image generation process. Layover always occurs at near range shadow always at far range of an object. That means, that data augmentation by rotation of SAR images would lead to nonsense imagery that would never be generated by a SAR.
    \vspace{1em}

    \item \textbf{The Complex Nature of SAR Data.} The most valuable information of SAR data lies in its phase. This applies for SAR image formation, which takes place in the complex signal domain, as well as for polarimetric, interferometric, and tomographic SAR data processing. This means that the entire CNN must be able to handle complex numbers. For the convolution operation this is trivial. The nonlinear activation function and the loss function, however, require thorough consideration. Depending on whether the activation function acts on the real and imaginary parts of the signal independently, or only on its magnitude, and where bias is added, phase will be distorted to different degrees.
    \par
    \vspace{0.5em}
    If we use polarimetric SAR data for land cover or target classification, a nonlinear processing of the phase is even desirable, because the phase between different polarimetric channels has physical meaning and, hence, contributes to the classification process.
    \par
    \vspace{0.5em}
    In SAR interferometry and tomography, however, the \textit{absolute} phase has no meaning, i.e., the CNN must be invariant to an arbitrary phase offset. Assume some interferometric input signal $x$ to a CNN and the output signal $CNN(x)$  with phase
    \begin{equation}
        \hat{\phi} = \angle CNN(x).
    \end{equation}
    Any constant phase offset $\phi_0$ does not change the meaning of the interferogram. Hence, we require an invariance that we refer to as "phase linearity" (valid at least in the expectation):
    \begin{equation}
        CNN(xe^{j\phi_0}) = CNN(x)e^{j\phi_0}.
    \end{equation}
    This linearity is violated, for example, if the activation function is applied to real and imaginary parts separately, or if a bias is added to the complex numbers.
    \par
    \vspace{0.5em}
    \hspace{2em}Another point to consider in regression-type InSAR CNN processing (e.g., for noise reduction) is the loss function. If the quantity of interest is not the complex number itself, but its phase, the loss function must be able to handle the cyclic nature of phases. It may also be advantageous that the loss function is independent\textemdash at least to a certain degree \textemdash of the signal magnitude to relieve the CNN from modelling the magnitude. A loss function that meets these requirements is, for example,
    \begin{equation}
        L=|E[e^{j(\angle CNN(x)-\angle y)}]|,
    \end{equation}
    where $y$ is the reference signal.
    \par
    \vspace{0.5em}
    \hspace{2em}Some authors use magnitude and phase, rather than real and imaginary parts, as input to the CNN. This approach is not invariant to phase offset, either. The interpretation of a phase function as a real-valued function forces the CNN to disregard the sharp discontinuities at the $\pm\pi$-transitions, whose positions are inconsequential. A standard CNN would pounce on these, interpreting them as edges.
     \vspace{1em}
 \item \textbf{Simulation-based Training and Validation Data?}     
The prevailing lack of ground-truth for regression-type tasks, like speckle reduction or InSAR denoising, might tempt us to use simulated SAR data for training and validation of neural networks. However, this bears the risk that our networks will learn models that are far too simplified. Unlike in the optical imaging field, where highly realistic scenes can be simulated, e.g. by PC games, the simulation of SAR data is more a scientific topic without the power of commercial companies and a huge market. SAR simulators focus on specific scenarios, e.g. vegetation (only distributed scatterers considered) or persistent (point) scatterers. The most advanced simulators are probably the ones for computing radar backscatter signatures of single military objects, like vessels. To our knowledge though there is no simulator available that can , e.g., generate realistic interferometric data of rugged terrain with layover, spatially varying coherence, and diverse scattering mechanisms. Often simplified scattering assumptions are made, e.g. that speckle is multiplicative. Even this is not true; pure Gaussian scattering can only be found for quite homogeneous surfaces and low resolution SARs. As soon as the resolution increases chances for a few dominating scatterers in a resolution cell increase as well and the statistics become substantially different from the one of fully developed speckle

    \end{itemize}
    
\section{Recent Advances in Deep learning applied to SAR}
\label{sec:app}

In this section, we provide an in-depth review of deep learning methods applied to SAR data from six perspectives: terrain surface classification, object detection, parameter inversion, despeckling, SAR Interferometry (InSAR), and SAR-optical data fusion. For each of these six applications, notable developments are stated in the chronological order, and their advantages and disadvantages are reported. Finally, each subsection is concluded with a brief summary. \SMrev{It is worth mentioning that the application of deep learning to SAR image formation is not explicitly treated here. For SAR focusing we have to distinguish between general-purpose focusing and imaging of objects with \textit{a priori} known properties, like sparsity. General-purpose algorithms produce data for applications like Land use and Land cover (LULC) classification, glacier monitoring, biomass estimation or interferometry. These are complex-valued focused data that retain all the information contained in the raw data. General-purpose focusing has a well-defined system model and requires a sequence of Fast Fourier Transform (FFT)s and phasor multiplications, i.e. linear operations like matrix-vector multiplications. For decades optimal algorithms have been developed to perform these operations at highest speed and with diffraction limited accuracy. There is no reason why deep neural networks should perform better or faster that this gold standard. If we want to introduce prior knowledge about the imaged objects, however, specialized focusing algorithms may be beneficially learned by neural networks. But even then it might make sense to focus raw data first by a standard algorithm and apply deep learning for post-processing. In \cite{gao_enhanced_2019} a CNN is trained to focus sparse military targets. But even in this approach the raw data are partially focused by FFT, before entering to the CNN.}

\subsection{Terrain Surface Classification}
\label{ssec:tsc}

As an important direction of SAR applications, terrain surface classification using PolSAR images is rapidly advancing with the help of deep learning. Regarding feature extraction, most conventional methods rely on exploring physical scattering properties \cite{moreira_tutorial_2013} and texture information \cite{he2013texture} in SAR images. However, these features are mainly human-designed based on specific problems and characteristics of data sources. Compared to conventional methods, deep learning is superior in terrain surface classification due to its capability of automatically learning discriminative features. Moreover, deep learning approaches, such as CNNs, can effectively extract not only polarimetric characteristics but also spatial patterns of PolSAR images \cite{parikh_classification_2020}. Some of the most notable deep learning techniques for PolSAR image classification are reviewed in the following.

Xie et al. \cite{xie2014multilayer} first applied deep learning to terrain surface classification using PolSAR images. They employed a stacked auto encoder (SAE) to automatically learn deep features from PolSAR data and then fed them to a softmax classifier. Remarkable improvements in both classification accuracy and visual effect proved that this method can effectively learn a comprehensive feature representation for classification purposes.

\begin{figure*}[!tbh]
\centering
\includegraphics[width=1\textwidth]{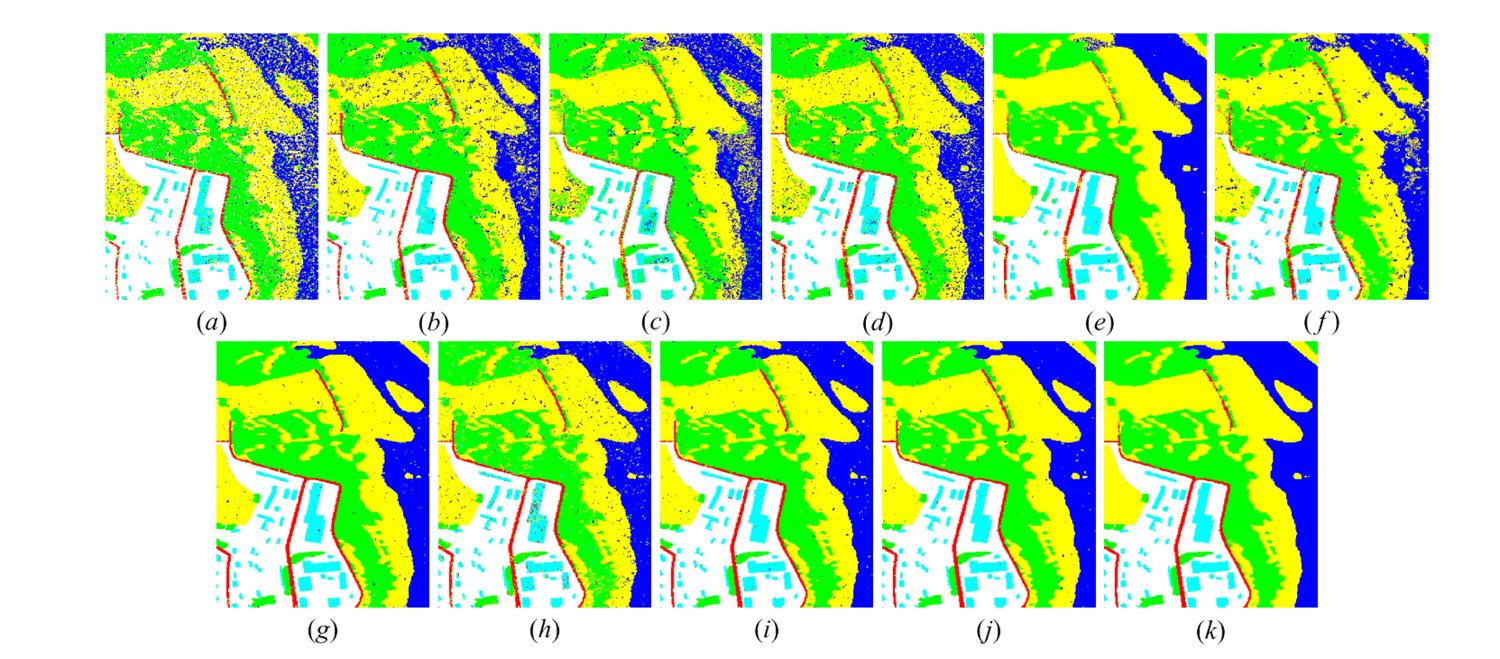}
\caption{Classification maps obtained from a TerraSAR-X image of a small area in Norway \cite{geng2017deep}. Subfigures (a)-(f) depict the results of classification using SVM (accuracy = 78.42\%), sparse representation classifier (SRC) (accuracy =  85.61\%), random forest (accuracy = 82.20\%) \cite{uhlmann2014integrating}, SAE (accuracy = 87.26\%) \cite{xie2014multilayer}, DCAE (accuracy = 94.57\%) \cite{geng2015high}, contractive AE (accuracy = 88.74).  Subfigures (g)-(i) show the combination of DSCNN with SVM (accuracy = 96.98\%), with SRC (accuracy = 92.51\%) \cite{hou_sar_2016}, and with random forest (accuracy = 96.87\%). Subfigures (j) and (k) represent the classification results of DSCNN (accuracy = 97.09\%) and DSCNN followed by spatial regularization (accuracy = 97.53\%), which achieve higher accuracy than the other methods.}
\label{fig:DSCNN}
\end{figure*}

Instead of simply applying SAE, Geng et al. \cite{geng2015high} proposed a deep convolutional autoencoder (DCAE) for automatically extracting features and performing classification. The first layer of DCAE is a hand-crafted convolutional layer, where filters are pre-defined, such as gray-level co-occurrence matrices and Gabor filters. The second layer of DCAE performs a scale transformation, which integrates correlated neighbor pixels to reduce speckle. Following these two hand-crafted layers, a trained SAE, which is similar to \cite{xie2014multilayer}, is attached for learning more abstract features. Tested on high-resolution single-polarization TerraSAR-X images, the method achieved remarkable classification accuracy.

Based on DCAE, Geng et al. \cite{geng2017deep} proposed a framework, called deep supervised and contractive neural network (DSCNN), for SAR image classification, which introduces histogram of oriented gradient (HOG) descriptors. In addition, a supervised penalty is designed to capture relevant information between features and labels, and a contractive restriction, which can enhance local invariance, is employed in the following trainable autoencoder layers. \SM{An example of applying DSCNN on TerraSAR-X data from a small area in Norway is seen in Fig.~\ref{fig:DSCNN}. Compared to other algorithms, the capability of DSCNN to achieve a highly accurate and noise free classification map is observed.}

\begin{figure*}[!ht]
\centering
\includegraphics[width=1\textwidth]{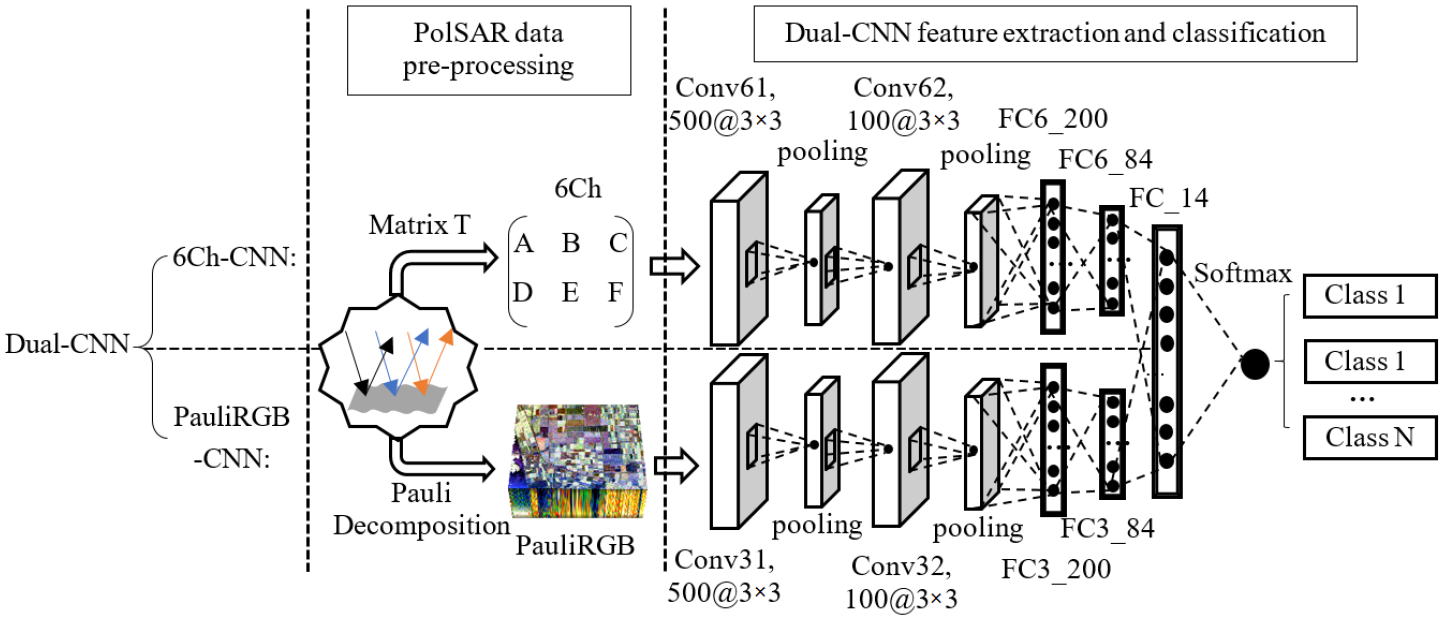}
\caption{The architecture of the dual-branch deep convolution neural network (Dual-CNN) for PolSAR image classification, proposed in \cite{gao2017dual}.}
\label{fig2.8}
\end{figure*}

In addition to the aforementioned methods, many studies integrate SAE models with conventional classification algorithms for terrain surface classification. Hou et al. \cite{hou2016classification} proposed an SAE combined with superpixel for PolSAR image classification. Multiple layers of the SAE are trained on a pixel-by-pixel basis. Superpixels are formed based on Pauli-decomposed pseudo-color images. Outputs of the SAE are used as features in the final step of k-nearest neighbor clustering of superpixels. Zhang et al. \cite{zhang2016stacked} applied stacked sparse AE to PolSAR image classification by taking into account local spatial information. Qin et al. \cite{qin2017object} applied adaptive boosting of RBMs to PolSAR image classification. Zhao et al. \cite{zhao2017discriminant} proposed a discriminant DBN (DisDBN) for SAR image classification, in which discriminant features are learned by combining ensemble learning with a deep belief network in an unsupervised manner. Moreover, taking into account that most current deep learning methods aim at exploiting features either from polarization information or spatial information of PolSAR images, Gao et al. \cite{gao2017dual} proposed a dual-branch CNN to learn features from both perspectives for terrain surface classification. This method is built on two feature extraction channels: one to extract polarization features from the 6-channel real matrix, and the other to extract spatial features of a Pauli decomposition. Next the extracted features are combined using two parallel fully connected layers, and finally fed to a softmax layer for classification. \SM{The detailed architecture of this network is illustrated in Fig. \ref{fig2.8}.}

Different variations of CNNs have been used for terrain surface classification as well. In \cite{zhou2016polarimetric}, Zhou et al. first extracted a 6-channel covariance matrix and then fed it to a trainable CNN for PolSAR image classification. Wang et al. \cite{wang2018hierarchical} proposed a fully convolutional network integrated with sparse and low-rank subspace representations for classifying PolSAR images. Chen et al. \cite{chen2018polsar} improved CNN performances by incorporating expert knowledge of target scattering mechanism interpretation and polarimetric feature mining. In a more recent work \cite{he_nonlinear_2020}, He et al. proposed the combination of features learned from nonlinear manifold embedding and applying a fully convolutional network (FCN) on input PolSAR images; the final classification was carried out in an ensemble approach by SVM. In \cite{dong_polsar_2020}, the authors focused on the computational efficiency of deep learning methods, proposing the use of lightweight 3D CNNs. They showed that classification accuracy comparable to other CNN methods was achievable while significantly reducing the number of learned parameters and therefore gaining computational efficiency. 

Apart from these single-image classification schemes using CNN, the use of time series of SAR images for crop classification has been shown in \cite{ndikumana_deep_2018, teimouri_novel_2019}. The authors of both papers experimented with using Recurrent Neural Network (RNN)-based architectures to exploit the temporal dependency of multi-temporal SAR images to improve classification accuracy. A unique approach for tackling PolSAR classification was recently proposed in \cite{dong_automatic_2020}, where for the first time the authors utilized an AutoML technique to find the optimum CNN architecture for each dataset. The approach takes into account the complex nature of PolSAR images, is cost effective, and achieves high classification accuracy \cite{dong_automatic_2020}.


Most of the aforementioned methods rely primarily on preprocessing or transforming raw complex-valued data into features in the real domain and then inputting them in a common CNN, which constrains the possibility of directly learning features from raw data. To tackle this problem, Zhang et al. \cite{zhang2017complex} proposed a novel complex-valued CNN (CV-CNN) specifically designed to process complex values in PolSAR data, i.e., the off-diagonal elements of a coherency or covariance matrix. The CV-CNN not only takes complex numbers as input but also employs complex weights and complex operations throughout different layers. A complex-valued backpropagation algorithm is also developed for CV-CNN training. Other notable complex-valued deep learning approaches for classification using PolSAR images can be found in \cite{mullissa_polsarnet_2019, li_complex_2019, xie_polsar_2020}. \SMrev{Different from the previously mentioned works, which exploit the complex-valued nature of SAR images in PolSAR image classification, Huang et al. \cite{huang_deep_2020} has recently proposed a novel deep learning framework called Deep SAR-Net for land use classification focusing on feature extraction from single-pol complex SAR images. The authors perform a feature fusion based on spatial features learned from intensity images and time-frequency features extracted from spectral analysis of complex SAR images. Since the time-frequency features are highly relevant for distinguishing different backscattering mechanisms within SAR images, they gain accuracy in classifying man-made objects compared to the use of typical CNNs, which only focus on spatial information.}

\SM{Although not completely related to terrain surface classification, it is also worth mentioning that the combination of SAR and PolSAR images with feed-forward neural networks has been extensively used for sea ice classification. This topic is not treated any further in this section and the interested reader is referred to consult \cite{ressel_neural_2015, ressel_neural_2016, ressel_investigation_2016, singha_arctic_2018, zakhvatkina_satellite_2019} for more information}. Similar to the polarimetric signature, InSAR coherence provides information about physical scattering properties. In \cite{mazza_tandem-x_2019} interferometric volume decorrelation is used as a feature for forest/non-forest mapping together with radar backscatter and incidence angle. The authors used bistatic TanDEM-X data where temporal decorrelation can be neglected. They compared different architectures and concluded that CNNs outperform random forest and U-Net \cite{ronneberger2015u} proved best for this segmentation task. 

To summarize, it is apparent that deep learning-based SAR and PolSAR classification algorithms have advanced considerably in the past few years. Although at first the focus was based on low-rank representation learning using SAE \cite{xie2014multilayer} and its modifications \cite{geng2015high}, later research focused on a multitude of issues relevant to SAR imagery, such as taking into account speckle \cite{geng2015high,geng2017deep} preserving spatial structures \cite{gao2017dual} and their complex nature \cite{zhang2017complex, mullissa_polsarnet_2019, li_complex_2019, huang_deep_2020}. It can also be seen that the challenge of the scarcity of labeled data has driven researchers to use semi-supervised learning algorithms \cite{xie_polsar_2020} \SMrev{although weakly supervised methods for semantic annotation, that has been proposed for high resolution optical data \cite{yao_semantic_2016}, has not been explicitly explored for classification tasks using SAR data. Furthermore, specific metric learning approaches to enhance class separability \cite{cheng_when_2018} can be adopted for SAR imagery in order to improve the overall classification accuracy.} Finally, one of machine learning's important fields, AutoML, a field that had not been exploited extensively by the remote sensing community, has found its application for PolSAR image classification \cite{dong_automatic_2020}.

\begin{figure*}[!ht]
\centering
\includegraphics[width=1\textwidth]{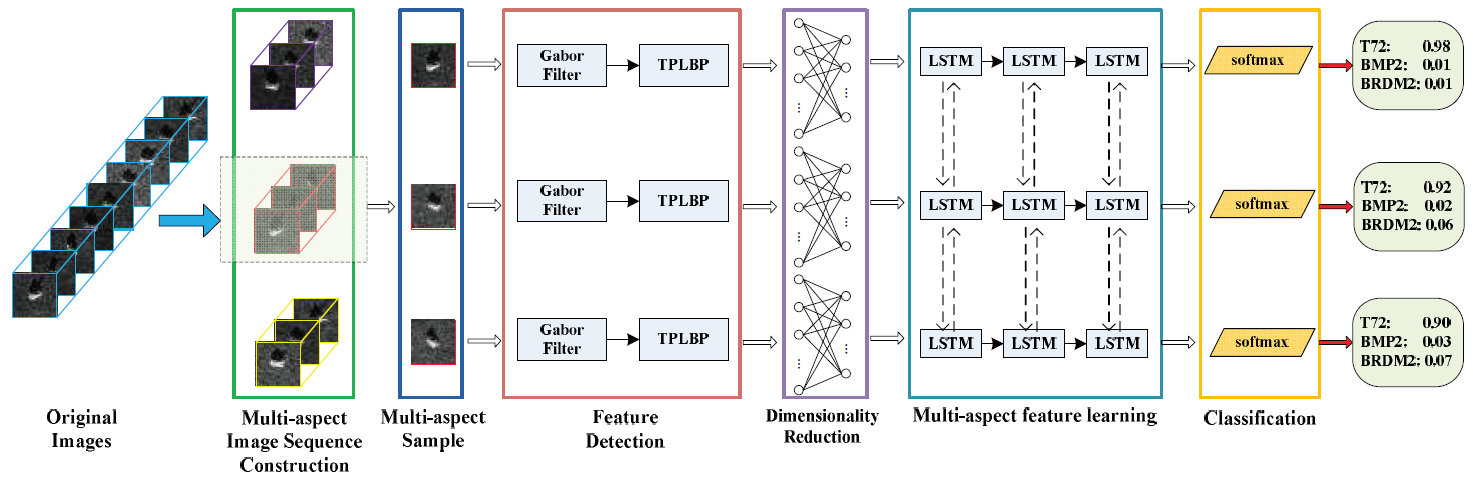}
\caption{The flowchart of the multi-aspect-aware bi-directional approach for SAR ATR proposed in \cite{zhang2017sar}.}
\label{fig2.7}
\end{figure*}

\subsection{Object Detection}
Although various characteristics distinguish SAR images from optical RGB images, the SAR object detection problem is still analogous to optical image classification and segmentation in the sense that feature extraction from raw data is always the  prior and crucial step. Hence, given success in the optical domain, there is no doubt that deep learning is one of the most promising ways to \YS{develop the state-of-the-art} SAR object detection algorithms.

The majority of earlier works on SAR object detection using deep learning consists of taking successful deep learning methods for optical object detection and applying them with minor tweaks to military vehicle detection (MSTAR dataset; see subsection \ref{ssec:mstar}) or ship detection on custom datasets. Even small-sized networks are easily able to achieve more than 90\% test accuracy on most of these tasks.

The first attempt in  military vehicle detection can be found in \cite{chen2014sar}, where Chen et al. used an unsupervised sparse autoencoder to generate convolution kernels from random patches of a given input for a single-layer CNN, which generated features to train a softmax classifier for classifying military targets in the MSTAR dataset \cite{keydel1996mstar}. The experiments in \cite{chen2014sar} showed great potential for applying CNNs to SAR target recognition. With this discovery, Chen et al. \cite{chen2016target} proposed A-ConvNets, a simple 5-layer CNN that was able to achieve state-of-the-art accuracy of about 99\% on the MSTAR dataset.

Following this trend, more and more authors applied CNNs to the MSTAR \YS{dataset} \cite{morgan2015deep, ding2016convolutional, du2016sar}. Morgan \cite{morgan2015deep} successfully applied a modestly sized 3-layered CNN on MSTAR and building upon it Wilmanski et al. \cite{micheal2016modern} investigated the effects of initialization and optimizer selection on final results. Ding et al. \cite{ding2016convolutional} investigated the capabilities of a CNN model combined with domain-specific data augmentation techniques (e.g., pose synthesis and speckle adding) in SAR object detection. Furthermore, Du et al. \cite{du2016sar} proposed a displacement- and rotation-insensitive CNN, and claimed that data augmentation on training samples is necessary and critical in the pre-processing stage. 


On the same dataset, instead of treating CNN as an end-to-end model, Wagner \cite{wagner2016sar} and similarly Gao \cite{gao2019new} integrated CNN and SVM, by first using a CNN to extract features, and then feeding them to an SVM for final prediction. Specifically, Gao et al. \cite{gao2017combining} added a class of separation information to the cross-entropy cost function as a regularization term, which they show explicitly facilitates intra-class compactness and separtability, in turn improving the quality of extracted features.
More recently, Furukawa \cite{furukawa2018deep}  proposed VersNet, an encoder-decoder style segmentation network, to not only identify but also localize multiple objects in an input SAR image. Moreover, Zhang et al. \cite{zhang2017sar} proposed an approach based on multi-aspect image sequences as a pre-processing step. In the contribution, they are taking into account  backscattering signals from different viewing geometries, following feature extraction using Gabor filters, dimensionallity reduction and eventually feeding the results to a Bidirectional LSTM model for joint recognition of targets. The flowchart of this SAR ATR framework is illustrated in Fig. \ref{fig2.7}.

Besides truck detection, ship detection is another tackled SAR object detection task. Early studies on applying deep learning models to ship detection \cite{cozzolino2017fully, schwegmann2016very, bentes2016ship, odegaard2016classification, liu2017sar} mainly consist of two stages: first cropping patches from the whole SAR image and then identifying whether cropped patches belong to target objects using a CNN. Because of fixed patch sizes these methods were not robust enough \MA{to cater for} variations in ship geometry, like size and shape. This problem was overcome by using region-based CNNs \cite{girshick2015fast, ren2017faster}, with creative use of skip connections and feature fusion techniques in later literature. For example, Li et al. \cite{li2017ship} fuses features of the last three convolution layers before feeding them to a region proposal network (RPN). Kang et al. \cite{kang2017contextual} proposed a contextual region based network that fuses features from different levels. Meanwhile, to make the most use of features of different resolution, Jiao et al. \cite{jiao2018densely} densely connected each layer to its subsequent layers and fed features from all layers to separate RPN to generat proposals; in the end the best proposal was chosen based on an intersection-over-union score.

In more recent works on SAR object detection, scientists have tried to explore many other interesting ideas to complement current works.  Dechesne et al. \cite{dechesne2019multi} proposed a multitask network that simultaneously learned to detect, classify, and estimate the length of ships.  Mullissa et al. \cite{mullissa2019polsarnet} showed that CNNs can be trained directly on Complex-Valued SAR data; Kazemi et al. \cite{kazemi2019deep} performed object classification using an RNN based architecture directly on received SAR signal instead of processed SAR images; and Rostami et al. \cite{rostami2019deep} and Huang et al. \cite{huang2019and} explored knowledge transfer or transfer learning from other domains to the SAR domain for SAR object detection.

\begin{figure*}
    \centering
    \includegraphics[width=1\textwidth]{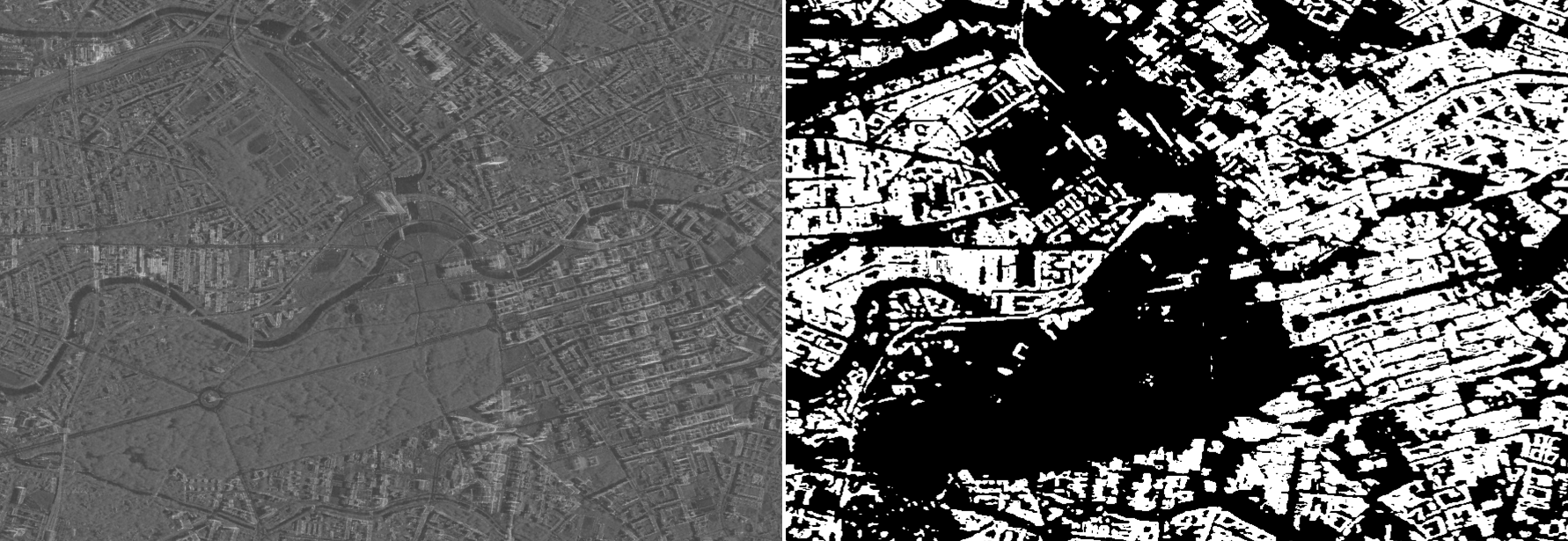}
    \caption{Very high resolution TerraSAR-X image of Berlin (left), and the predicted building mask \cite{shahzad2019buildings} (right).}
    \label{fig:shahzad_building}
\end{figure*}

Perhaps one of the more interesting recent works in this application area is building detection by Shahzad et al. \cite{shahzad2019buildings}. They tackle the problem of Very High Resolution (VHR) SAR building detection using a \YS{FCN} \cite{long2015fully} architecture for feature extraction, followed by CRF-RNN \cite{zheng2015conditional}, which helps give similar weights to neighboring pixels. This architecture produced building segmentation masks with up to 93\% accuracy. \YW{An example of the detected buildings can be seen in Fig. \ref{fig:shahzad_building}, where the left subfigure is the amplitude of the input TerraSAR-X image of Berlin, and the right subfigure is the predicted building mask.} Another major contribution made in \YS{that} paper addresses the problem of lack of training data by introducing an automatic annotation technique, which annotates the TomoSAR data using Open Street Map (OSM) data.

As an extension of the abovementioned work, Sun et al. \cite{sun2020cg} tackled the problem of individual building segmentation in large-scale urban areas. They propose a conditional GIS-aware network (CG-Net) that learns multi-level visual features and employs building footprint data to normalize these features for predicting building masks. Thanks to the novel network architecture and the large amounts buildings labels automatically generated from an accurate DEM and GIS building footprints, this network achieves F1 score of 75.08\% for individual building segmentation. With the predicted building masks, large-scaled levels-of-detail (LoD) 1 building models are reconstructed with mean height error of 2.39 m. 

\MAN{Overall deep learning has shown very good performance in existing SAR object detection tasks. There are two main challenges that the algorithm designer needs to keep in mind when tackling any SAR object detection tasks.} The first is the challenge of identifying characteristics of SAR imagery like imaging geometry, size of objects, and speckle noise. The second and bigger challenge is the \YS{lack} of good quality standardized datasets. As we observed, the most popular dataset, MSTAR, is too easy for deep nets, and for ship detection, the majority of authors created their datasets, which makes it very hard to judge the quality of the proposed algorithms and even harder to compare different algorithms. \MAN{An example of difficult to create dataset is that of a dataset for global building detection. The shape, size, and style of the buildings changes from region to region quite drastically, and so a good dataset for this purpose requires training examples of buildings from around the world which needs quite a big effort to do high quality annotation of enough number of buildings such that deep nets can learn something from them.}
\subsection{Parameter Inversion}
\label{ssec:ice}

Parameter inversion from SAR images is a challenging field in SAR applications. As one important branch, ice concentration estimation is now attracting great attention due to its importance to ice monitoring and climate research \cite{radar2014global}. Since there are complex interactions between SAR signals and sea ice \cite{dierking2013sea}, empirical algorithms face difficulties with interpreting SAR images for accurate ice concentration estimation. 

Wang et al. \cite{wang2015ice} resorted to a CNN for generating ice concentration maps from dual polarized SAR images. Their method takes image patches of the intensity-scaled dual band SAR images as inputs, and outputs ice concentration directly. In \cite{wang2016sea,wang2016learning}, Wang et al. employed various CNN models to estimate ice concentration from SAR images during the melt season. Labels are produced by ice experts via visual interpretation. The algorithm was tested on dual-pol RadarSat-2 data. Since the problem considered is the regression of a continuous value, mean squared error is selected as the loss function. Experimental results demonstrate that CNNs can offer a more accurate result than comparative operational products.

\begin{figure*}[!ht]
\centering
\includegraphics[width=1\textwidth]{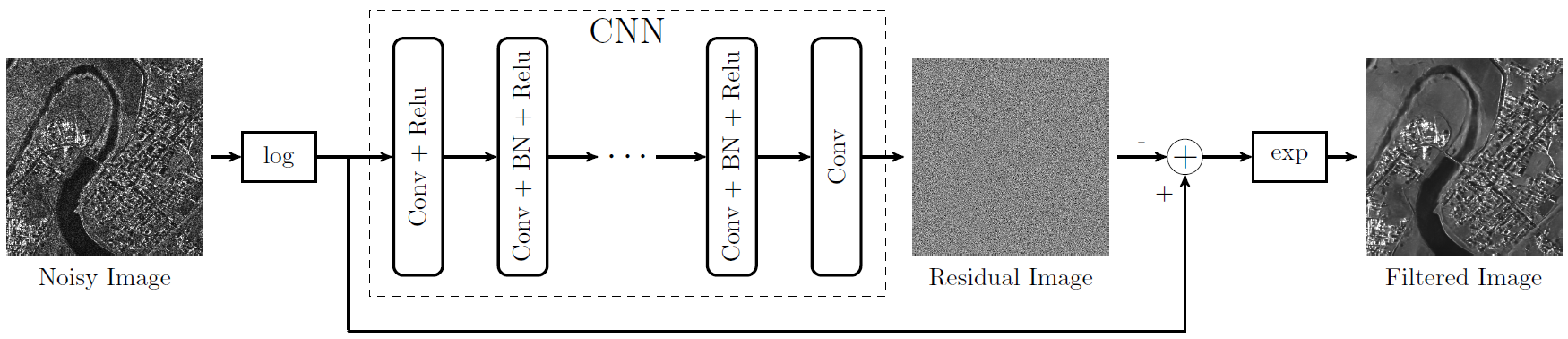}
\caption{The Architecture of CNN for SAR image despeckling \cite{chierchia2017sar}.}
\label{fig2.6}
\end{figure*}

\begin{figure*}[t!h]
\centering
\includegraphics[width=.8\textwidth]{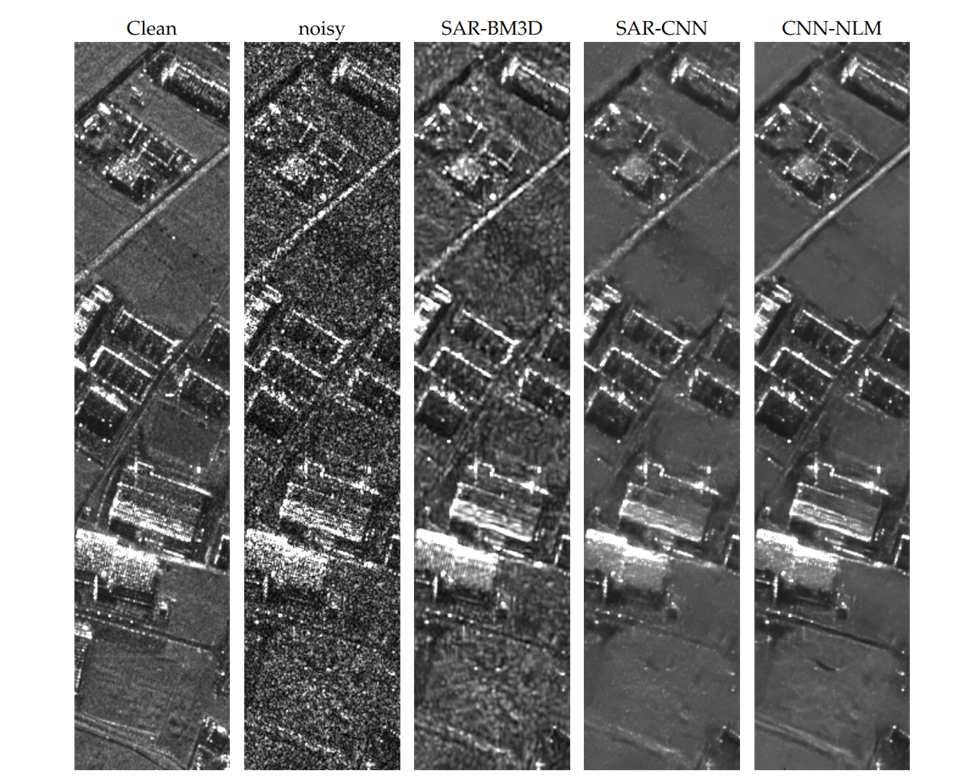}
\caption{The comparison of speckle reduction among SAR-BM3D \cite{parrilli_nonlocal_2012}, SAR-CNN \cite{chierchia2017sar}, and CNN-NLM applied to a small strip of COSMO-SkyMed data over Caserta, Italy, where the reference clean image has been obtained by temporal multi-looking applied to a stack of SAR images \cite{cozzolino_nonlocal_2020}.}
\label{fig:CNN_NLM}
\end{figure*}

In a different application, Song et al. \cite{song_inversion_2018} used a deep CNN, including five pairs of convolutional and max pooling layers followed by two fully connected layers for inverting rough surface parameters from SAR images. The training of the network was based on simulated data solely due to the scarcity of real training data. The method was able to invert the desired parameters with a reasonable accuracy and the authors showed that training a CNN for parameter inversion purposes could be done quite efficiently.  Furthermore, Zhao et al. \cite{zhao2019contrastive} designed a complex-valued CNN to directly learn physical scattering signatures from PolSAR images. The authors have notably proposed a framework to automatically generate labeled data, which led to a supervised learning algorithm for the aforementioned parameter inversion. \SMrev{The approach is similar to the study presented in \cite{song_radar_2018}, where the authors used deep learning for SAR image colorization and learning a full polarimteric SAR image from single-pol data. Another interesting application of deep learning in parameter inversion has been recently published in \cite{niu_parameter_2020}. The authors propose a deep neural network architecture containing a CNN and a GAN to automatically learn SAR image simulation parameters from a small number of real SAR images. They later feed the learned parameters to a SAR simulator such as RaySAR \cite{auer_raysar_2016} to generate a wide variety of simulated SAR images, which can increase training data production and improve the interpretation of SAR images with complex backscattering scenarios.}

On the whole, deep learning-based parameter estimation for SAR applications has not yet been fully exploited. Unfortunately, most of the focus of the remote sensing community has been devoted to classical problems, which overlap with computer vision tasks such as classification, object detection, segmentation, and denoising. \SMrev{One reason for this might be since parameter estimation usually requires the incorporation of appropriate physical models and tackles the problem at hand as regression rather than classification, the domain knowledge is quite essential in order to apply deep learning for such tasks, especially for SAR images with their peculiar physical characteristics. One interesting study \cite{huang_deep_2020} that has been already described in detail in subsection \ref{ssec:tsc}, which designs discriminative features by spectral analysis of complex-valued SAR data is an important work toward including deep learning in parameter inversion studies using SAR data.}
We hope that in the future more studies will be carried out in this direction.

\subsection{Despeckling}
\label{sec:despeck}

Speckle, caused by the coherent interaction among scattered signals from sub-resolution objects, often makes processing and interpretation of SAR images difficult. Therefore, despeckling is a crucial procedure before applying SAR images to various tasks. Conventional methods aim at removing speckle either spatially, where local spatial filters, such as the Lee filter \cite{lee1980digital}, Kuan filter \cite{kuan1985adaptive}, and Frost filter \cite{frost1982model}, are employed, or using wavelet-based methods \cite{xie2002sar, argenti2002speckle, achim2003sar}. For a full overview of these techniques, the reader is referred to \cite{argenti_tutorial_2013}. In the past decade, patch-based methods for speckle reduction have gained high popularity due to their ability to preserve spatial features while not sacrificing image resolution \cite{tupin_ten_2019}. Deledalle et al. \cite{deledalle_iterative_2009} proposed one of the first nonlocal patch-based methods applied to speckle reduction by taking into account the statistical properties of speckle combined with the original nonlocal image denoising algorithm introduced in \cite{buades_non-local_2005}. A vast number of variations of the nonlocal method for SAR despeckling has been proposed, with the most notable ones included in \cite{xin_su_two-step_2014, deledalle_nl-sar_2015}. However, on one hand, manual selection of appropriate parameters for conventional algorithms is not easy and is sensitive to reference images. On the other hand, it is difficult to achieve a balance between preserving distinct image features and \SM{removing} artifacts with empirical despeckling methods. To solve these limitations, methods based on deep learning have been developed.

Inspired by the success of image denoising using a residual learning network architecture in the computer vision community \cite{zhang2017beyond}, Chierchia et al. \cite{chierchia2017sar} first introduced a residual learning CNN for SAR image despeckling by presenting a 17-layered CNN for learning to subtract speckle components from noisy images. Considering that speckle noise is assumed to be multiplicative, the homomorphic approach with coupled log- and exp-transformations is performed before and after feeding images to the network. In this case, multiplicative speckle noise is transformed into an additive form and can be recovered by residual learning, where log-speckle noise is regarded as residual. As shown in Fig. \ref{fig2.6}, an input log-noisy image is mapped identically to a fusion layer via a shortcut connection, and then added element-wise with the learned residual image to produce a log-clean image. Afterwards, denoised images can be obtained by an exp-transformation. Wang et al. \cite{wang2017sar} proposed a CNN, called ID-CNN, for image despeckling, which can directly learn denoised images via a component-wise division-residual layer with skip connections. In another words, homomorphic processing is not introduced for transforming multiplicative noise into additive noise and at a final stage the noisy image is divided by the learned noise to yield the clean image.

As a step forward with respect to the two aforementioned residual-based learning methods, Zhang et al. \cite{zhang2018learning} employed a dilated residual network, SAR-DRN, instead of simply stacking convolutional layers. Unlike \cite{chierchia2017sar} and similar to \cite{wang2017sar}, SAR-DRN is trained in an end-to-end fashion using a combination of dilated convolutions and skip connections with a residual learning structure, which indicates that prior knowledge such as a noise description model is not required in the workflow.

In \cite{yue2018sar}, Yue et al. proposed a novel deep neural network architecture specifically designed for SAR despeckling. It used a convolutional neural network to extract image features and reconstruct a discrete RCS probability density function (PDF). It is trained by a hybrid loss function which measures the distance between the actual SAR image intensity PDF and the estimated one that is derived from convolution between the reconstructed RCS PDF and prior speckle PDF. Experimental results demonstrated that the proposed despeckling neural network can achieve comparable performance as non-learning state-of-the-art methods. 

\YW{The unique distribution of SAR intensity images was also taken into account in \cite{vitale_multi-objective_2020}. It proposed a different loss function which contains three terms between the true and the reconstructed image. They are the common L2 loss, the L2 difference between the gradient of the two images, and the Kullback-Leibler divergence between the distribution of the two images. The three terms are designed to emphasize the spatial details, the identification of strong scatterers, and the speckle statistics, respectively. Experiments in \cite{vitale_multi-objective_2020} show improved performance compared to SAR-BM3D \cite{parrilli_nonlocal_2012} and SAR-DRN \cite{zhang2018learning}.}

In \cite{tang2018sar}, the problem of despeckling was tackled by a time series of images. Using a stack of images for despeckling is not unique to deep learning-based methods, as has been recently demonstrated in \cite{baier_robust_2020} as well. In \cite{tang2018sar} the authors utilized a multi-layer perceptron with several hidden layers to learn non-linear intensity characteristics of training image patches. This approach has shown promising results and reported comparative performance with the state-of-the-art despeckling algorithms.

Again using single images instead of time series, in \cite{lattari_deep_2019} the authors proposed a deep encoder–decoder CNN architecture with focus on feature preservation, which is a weakness of CNNs. They modified U-Net \cite{ronneberger2015u} in order to accommodate speckle statistical features. Another notable CNN approach was introduced in \cite{cozzolino_nonlocal_2020}, where the authors used a nonlocal structure, while the weights for pixel-wise similarity measures were assigned using a CNN. The results of this approach, called CNN-NLM, are reported in Fig.~\ref{fig:CNN_NLM}, where the superiority of the method with respect to both feature preservation and speckle reduction is clearly observed. 

\YW{One of the drawbacks of the aforementioned algorithms is the requirement of noise-free and noisy image pairs for training. Often, those training data are simulated using optical images with multiplicative noise. This is of course not ideal for real SAR images. Therefore, one elegant solution is the noise2noise framework \cite{lehtinen2018noise2noise}, where the network only requires two noisy images of the same area. \cite{lehtinen2018noise2noise} proves that the network is able to learn a clean representation of the image given the noise distributions of the two noisy images are independent and identical. This idea has been employed in SAR despeckling in \cite{ma_sar_2020}. The authors make use of multi-temporal SAR images of a same area as the input to the noise2noise network. To mitigate the effect of the temporal change between the input SAR image pairs, the authors multiples a patch similarity term to the original loss function.}

From the deep learning-based despeckling methods reviewed in this subsection, it can be observed that most methods employ CNN-based architectures with single images of the scene for training; they either output the clean image in an end-to-end fashion or propose residual-based techniques to learn the underlying noise model. With the availability of large archives of time series thanks to the Sentinel-1 mission, an interesting direction is to exploit the temporal correlation of speckle characteristics for despeckling applications. \SMrev{One critical issue is over-smoothing in despeckling that needs to be addressed. Many of the CNN-based methods perform well in terms of speckle removal but are not able to preserve sharp edges. This is quite problematic in despeckling of high resolution SAR images of urban areas in particular.} Another problem in supervised deep learning-based despeckling techniques is the lack of ground truth data. In many studies, the training data set is built by corrupting optical images by multiplicative noise. This is far from realistic for despeckling applied to real SAR data. Therefore, despeckling in an unsupervised manner would be highly desirable and worth attention.

\subsection{InSAR}
\label{sec:insar}

Interferometric SAR (InSAR) is one of the most important SAR techniques, and is widely used in reconstructing the topography of the Earth's surface, i.e., digital elevation model (DEM) generation \cite{zebker1994accuracy, abdelfattah2002topographic, moreira_tutorial_2013}, and detecting topographical displacements, e.g., monitoring volcanic eruptions \cite{massonnet1995deflation, ruch2008caldera, trasatti20082004}, earthquakes \cite{massonnet1993displacement, peltzer1995surface}, land subsidence \cite{gini_ketelaar_satellite_2009}, and urban areas using time series methods \cite{zhu_lets_2011, gernhardt_deformation_2012, montazeri_three-dimensional_2016}. 

The principle of InSAR is to first measure the interferometric phase between signals received by two antennas located at different positions and then extract topographic information from the obtained interferogram by unwrapping and converting the absolute phase to height. However, an actual interferogram often suffers from a large number of singular points, which originate from the interference distortion and noise in radar measurements. These points result in unwrapping errors and consequently low quality DEMs. To tackle this problem, Ichikawa and Hirose \cite{ichikawa2017singular} applied a complex-valued neural network, CVNN, in the spectral domain to restore singular points. With the help of the Complex Markov Random Field (CMRF) filter \cite{yamaki_singular_2009}, they aimed at learning ideal relationships between the spectrum of neighboring pixels and that of center pixels via a one-hidden-layer CVNN. Notably, center pixels of each training sample are supposed to be ideal points, which indicate that singular points are not fed to the network during the training procedure. Similarly, Oyama and Hirose \cite{oyama2018adaptive} restored singular points with a CVNN in the spectrum domain.

\begin{figure*}[t!hb]
\centering
\includegraphics[width=.9\textwidth]{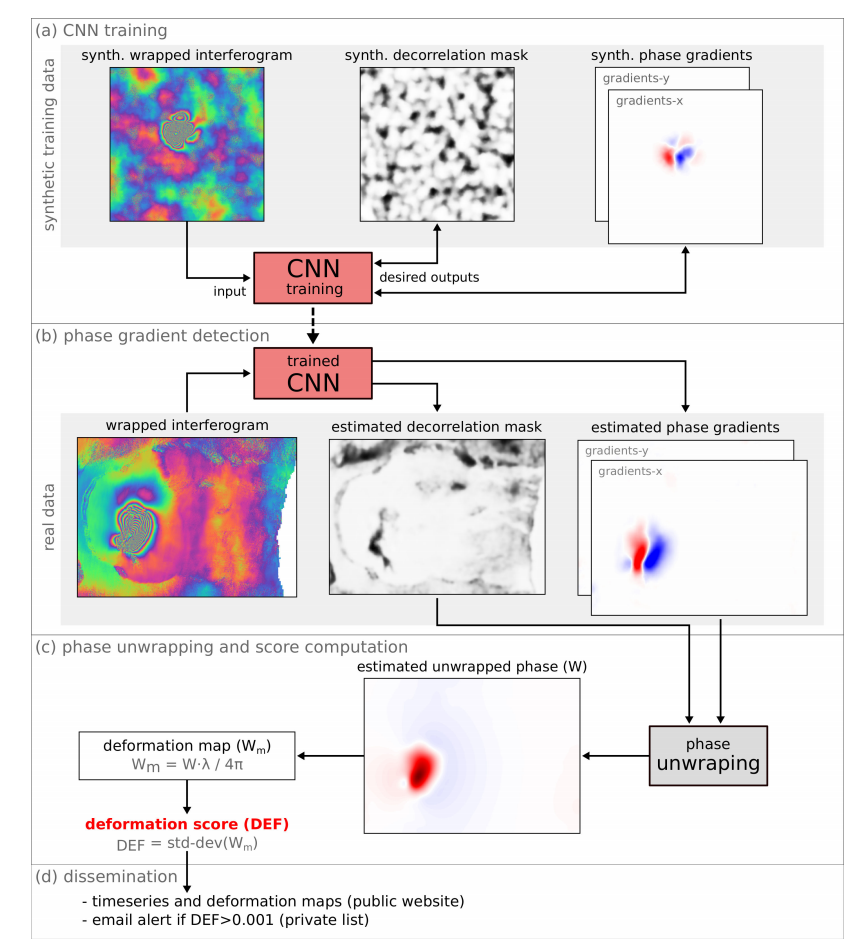}
\caption{The workflow of volcano deformation detection proposed in \cite{valade_towards_2019}. The CNN is trained on simulated data and is later used to detect phase gradients and a decorrelation mask from input wrapped interferograms to locate ground deformation caused by volcanoes.}
\label{fig:volcano_detection}
\end{figure*}

Related to topography extraction, Costante et al. \cite{costante2018towards} proposed a fully CNN Encoder-Decoder architecture for estimating DEM from single-pass image acquisitions. It is demonstrated that this model is capable of extracting high-level features from input radar images using an encoder section and then reconstructing full resolution DEM via a decoder section. Moreover, the network can potentially solve the layover phenomenon in one single-look SAR image with contextual features.

In addition to reconstructing DEMs, Schwegmann et al. \cite{schwegmann2017subsidence} presented a CNN-based technique to detect subsidence deformations from interferograms. They employed a 9-layer network to extract salient information in interferograms and displacement maps for discriminating deformation targets from deformation-like targets. Furthermore, Anantrasirichai et al. \cite{anantrasirichai_application_2018,anantrasirichai2018detecting, anantrasirichai_deep_2019} used a pre-trained CNN to automatically detect volcanic ground deformation from InSAR images. They divided each image into patches, and relabeled them with binary labels, i.e., $"$background$"$ and $"$volcano$"$, and finally fed them to the network to predict volcano deformation. They further improved their method to be able to detect slow-moving volcanoes by using a time series of interferograms in \cite{anantrasirichai_application_2019}. In another study related to automatic volcanic deformation detection, Valade et al. \cite{valade_towards_2019} designed and trained a CNN from scratch to learn a decorrelation mask from input wrapped interferograms, which then was used to detect volcanic ground deformation. The flowchart of this approach can be seen in Fig.~\ref{fig:volcano_detection}. The training in both of the aforementioned works \cite{anantrasirichai_application_2019, valade_towards_2019} was based on simulated data. Another geophysically motivated example of using deep learning on InSAR data, which was actually proposed earlier than the above-mentioned CNN-based studies, was seen in \cite{del_frate_neural_2010, picchiani_neural_2011, stramondo_seismic_2011}, where the authors used simple feed-forward shallow neural networks for seismic event characterization and automatic seismic source parameter inversion by exploiting the power of neural networks in solving non-linear problems.

\begin{figure*}[!th]
\centering
\includegraphics[width=1\textwidth]{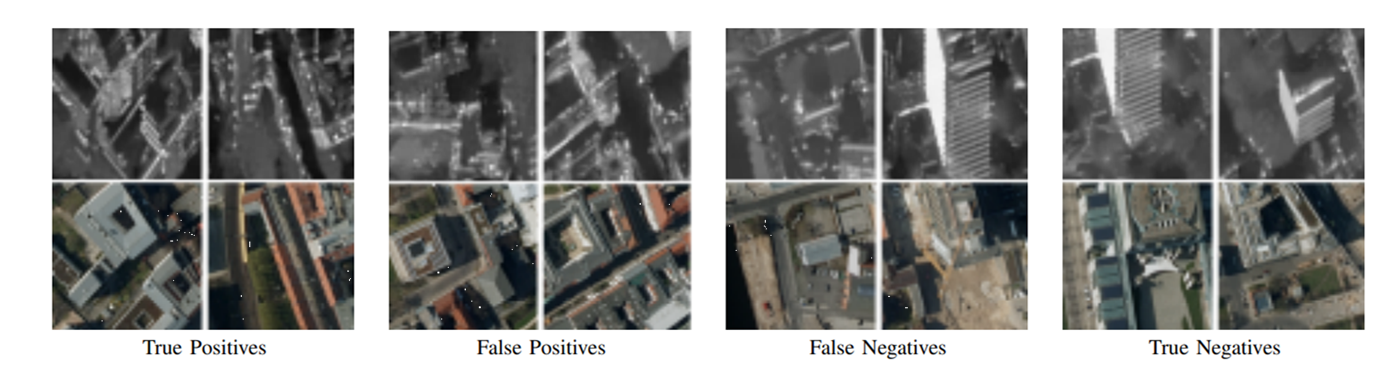}
\caption{Randomly selected patches obtained from the testing phase of the network for SAR-optical image patch correspondence detection proposed in \cite{hughes2018identifying}.}
\label{fig:SAR-opt_patches}
\end{figure*}

\YS{Recently, deep learning has been utilized for tomographic processing as well. An unfolded deep network which involves the vector approximate message passing algorithms has been proposed in \cite{gao2019fast}. Experiments with simulated and real data have been performed, which shows the spectral estimation gains speed up and achieves competitive performance. In \cite{wu2020super}, a real-valued deep neural network is applied for MIMO SAR 3-D imaging. It shows a better super-resolution power compared with other compressive sensing-based methods.}

In summary, it can be concluded that the use of deep learning methods in InSAR is still at a very early stage. Although deep learning has been used in different applications combined with InSAR, the full potential of interferograms is not yet fully exploited except in the pioneering work of Hirose \cite{hirose_complex-valued_2012}. Many applications treat interferograms or deformation maps obtained from interferograms as images similar to RGB or gray-scale ones and therefore the complex nature of interferograms has remained unnoticed. Apart from this issue, like the SAR despeckling problem using deep learning, lack of ground truth data for either detection or image restoration problems is a motivation to focus on developing semi-supervised and unsupervised algorithms that combine deep learning and InSAR. \SMrev{Otherwise a training database consisting of interferograms for different scenarios and also for different phase contributions could be beneficial for supervised learning applications. Simulation-based interferogram generation for the latter has been recently proposed \cite{rongier_generative_2019}.}


\subsection{SAR-Optical Data fusion}
\label{sec:patch}

The fusion of SAR and optical images can provide complementary information about targets. However, considering the two different sensing modalities, prior identification and co-registration of corresponding images are challenging \cite{schmitt2016challenges}, but compulsory for joint applications of SAR and optical images. For the purpose of identifying and matching SAR and optical images, many current methods resort to deep learning, given its powerful capabilities of extracting effective features from complex images.

In \cite{mou2017cnn}, the authors proposed a CNN for identifying corresponding image patches of very high resolution (VHR) optical and SAR imagery of complex urban scenes. Their network consists of two streams: one designed for extracting features from optical images, the other responsible for learning features from SAR images. Next the extracted features are fused via a concatenation layer for further binary prediction of their correspondence. A selection of True Positives, False Positives, False Negatives, and True Negatives of SAR-optical image patches from \cite{mou2017cnn} can be seen in Fig.~\ref{fig:SAR-opt_patches}. Similarly, Hughes et al. \cite{hughes2018identifying} proposed a pseudo-Siamese CNN for learning a multi-sensor correspondence predictor for SAR and optical image patches. Notably, both networks in \cite{mou2017cnn, hughes2018identifying} are trained and validated on the SARptical dataset \cite{wang2017potential, wang2018sarptical}, which is specifically built for joint analysis of VHR SAR and optical images in dense urban areas.

In \cite{wang2018deep}, the authors proposed a deep learning framework that can learn an end-to-end mapping between image patch pairs and their matching labels. An image pair is first transformed into two 1-D vectors and then concatenated to build a large 1-D vector as the input of the network. Then hidden layers are stacked for learning the mapping between input vectors and output binary labels, which indicate their correspondence.

For the purpose of matching SAR and optical images, Merkle et al. \cite{merkle2017exploiting} presented a CNN that comprises of a feature extraction stage (Siamese network) and a similarity measure stage (dot product layer). Specifically, features of input optical and SAR images are extracted via two separate 9-layer branches and then fed to a dot product layer for predicting the shift of the optical image within the large SAR reference patch. Experimental results indicate that this deep learning-based method outperforms state-of-the-art matching approaches \cite{suri2010mutual,dellinger_sar-sift_2015}. Furthermore, Abulkhanov et al. \cite{abulkhanov2018neural} successfully trained a neural network to build feature point descriptors to identify corresponding patches among SAR and optical images and match the detected descriptors using the RANSAC algorithm  \cite{fischler_random_1981}. 

In contrast to training a model to identify corresponding image patches, Merkle et al. \cite{merkle2018exploring} first employed a conditional generative adversarial network (cGAN) to generate artificial SAR-like images from optical images, then matched them with real SAR images. The authors demonstrate that the matching accuracy and precision are both improved with the proposed strategy. Inspired by their study, more researchers resorted to using GANs for the purpose of SAR-optical image matching (see \cite{hughes_deep_2019, fuentes_reyes_sar--optical_2019} for a review).


With respect to applications of SAR and optical image matching, Yao et al. \cite{yao2017semantic} aimed at applying SAR and optical images to semantic segmentation with deep neural networks. They collected corresponding optical patches from Google Earth according to TerraSAR-X patches and built ground truths using data from OpenStreetMap. Then SAR and optical images were separately fed to different CNNs to predict semantic labels (building, natural, land use, and water). \YW{Despite their experimental} results \YW{not outperforming the state of the art by the time \cite{lai_semantic_2017} likely because of network design or training strategy}, they deduced that introducing advanced models and simultaneously using both data sources can greatly improve the performance of semantic segmentation. Another application mentioned in \cite{schmitt2018colorizing} demonstrated that standard fusion techniques for SAR and optical images require data from both sources, which indicates that it is still not easy to interpret SAR images without the support of optical images. To address this issue, Schmitt et al. \cite{schmitt2018colorizing} proposed an automatic colorization network, composed of a VAE and a mixture density network (MDN) \cite{bishop1994mixture}, to predict artificially colored SAR images (i.e., Sentinel-1 images). These images are proven to disclose more information to the human interpreter than the original SAR data.

In \cite{grohnfeldi2018conditional}, the authors tackled the problem of cloud removal from optical imagery. They introduced a cGAN architecture to fuse SAR and cloud-corrupted multi-spectral data for generating cloud- and haze-free multi-spectral optical data. Experiments proved the effectiveness of the proposed network for removing cloud from multi-spectral data with auxiliary SAR data. Extending previous multi-modal networks for cloud removal, \cite{Ebel20} proposed a cycle-consistent GAN architecture \cite{Zhu_Park_Isola_Efros_2017} that utilizes a \YW{image forward-backward translation consistency} loss. Cloud-covered optical information is reconstructed via SAR data fusion, while changes to cloud-free areas are minimized through use of the cycle consistency loss. The cycle-consistent architecture allows training without pixel-wise correspondences between cloudy input and cloud-free target optical imagery, relaxing requirements on the training data set.

In summary, it can be seen that the utilization of deep learning methods for SAR-optical data fusion has been a hot topic in the remote sensing community. Although a handful of data sets consisting of optical and SAR corresponding image patches are available for different terrain types and applications, one of the biggest problems in this task is still the scarcity of high quality training data. Semi-supervised methods, as proposed in \cite{hughes_semi-supervised_2019}, seems to be a viable option to tackle the problem. A great challenge in the SAR-optical image matching is the extreme difference in viewing geometries of the two sensors. For this it is important to exploit auxiliary 3D data in order to assist the training data generation.

\begin{figure*}[!ht]
\centering
\includegraphics[width=1\textwidth]{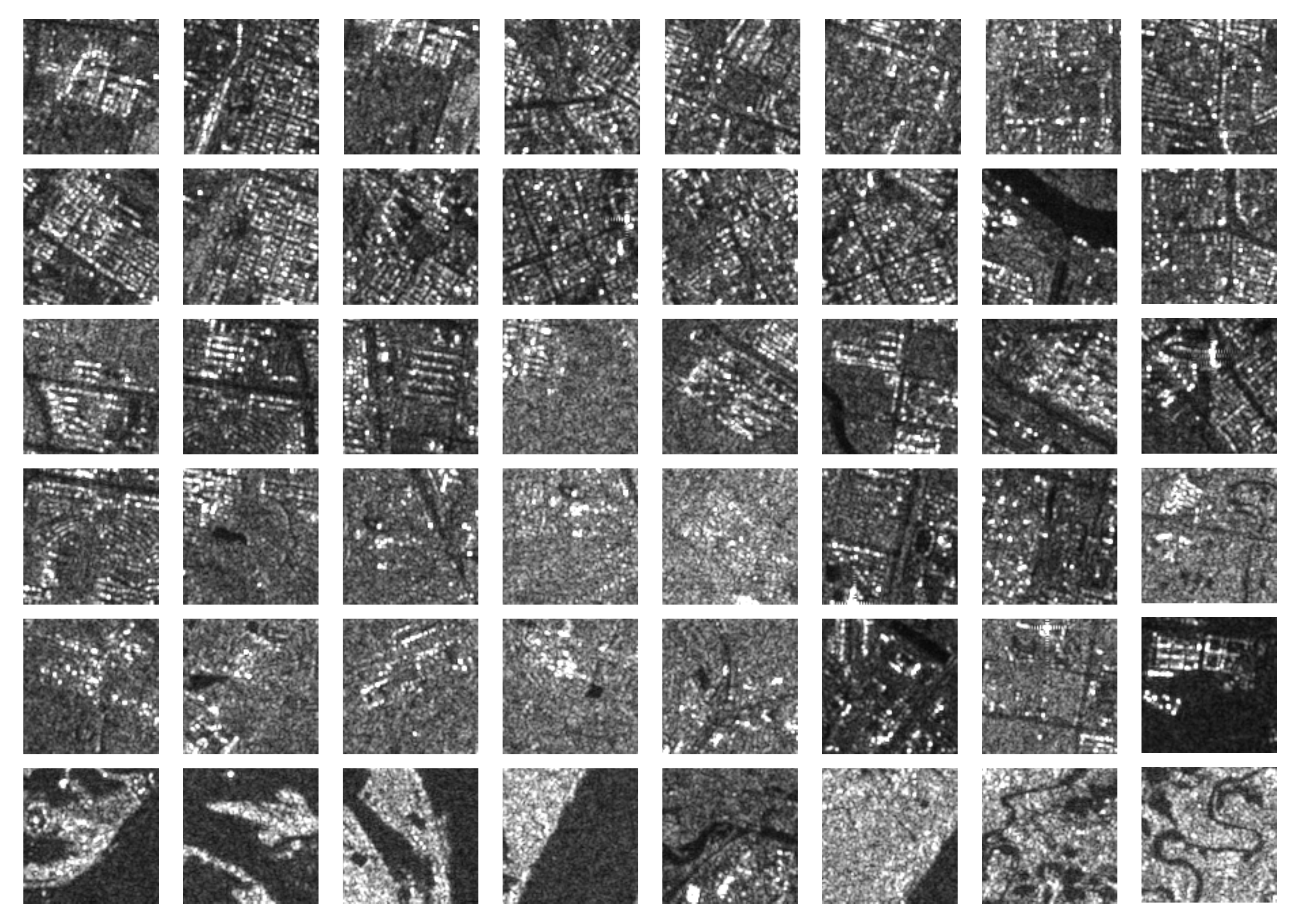}
\caption{Samples of the OpenSARUrban  \cite{zhao_opensarurban_2020}. Six classes are shown from the top to the bottom: dense and low-rise residential buildings, general residential area, high-rise buildings, villas, industrial storage area, and vegetation.}
\label{fig_opensarurban}
\end{figure*}

\section{Existing Benchmark Datasets and their limitations}
\label{sec:data}
In order to train and evaluate deep learning models, large datasets are indispensable. Unlike RGB images in the computer vision community, which can be easily collected and interpreted, SAR images are much more difficult to annotate due to their complex properties. Our research shows that big SAR datasets created for the primary purpose of deep learning research are nearly non-existent in the community. In recent years, only a few SAR datasets have been made public for training and assessing deep learning models. In the following, we categorize those datasets according to their best suited deep learning problem and focus on openly accessible and well-curated large datasets.   


In particular, we consider the following categories of deep learning problems in SAR.
\begin{itemize}
    \item Image classification: each pixel or patch in one image is classified into a single label. This is often the case in typical land use land cover classification problems.
    \item Scene classification: similar to image classification, one image or patch is classified into a single label. However, one scene is usually much larger than an image patch. Hence, it requires a different network architecture.
    \item Semantic segmentation: one image or patch is segmented to a classification map of the same dimension. Training of such neural networks also requires densely annotated training data.
    \item Object detection: similar to scene classification. However, detection often requires the estimation of the object \YW{location}. 
    \item Registration/matching: provide binary classification (matched or unmatched), or estimate the translation between two image patches. This type of task requires matching pairs of two different image patches as training data.
\end{itemize}

\begin{table*}[ht]
\label{tab:summary_datasets}
\centering

\caption{Summary of available open SAR datasets}
\begin{tabular}{p{3cm}p{9cm}p{2.7cm}p{1.8cm}}
\toprule[1.0pt]
Name & Description & Suitable tasks & \ Related work \\
\midrule

So2Sat LCZ42\footnotemark ~\cite{zhu2019so2sat}, \newline TensorFlow API\footnotemark  & 400,673 pairs of corresponding Sentinel-1 dual-pol image patch, Sentinel-2 multispectral image patch, and manually labeled local climate zones classes over 42 urban agglomerations (plus 10 additional smaller areas) across the globe. It is the first EO dataset that provides a quantitative measure of the label uncertainty, achieved by having a group of domain experts cast 10 independent votes on 19 cities in the dataset. & image classification, \newline data fusion, \newline quantification of uncertainties &  \cite{neumann_-domain_2019}\\

OpenSARUrban\footnotemark ~\cite{zhao_opensarurban_2020}& 33,358 Sentinel-1 dual-pol images patches covering 21 major cities in China, labeled with 10 classes of urban scenes. & image classification & \\

SEN12MS\footnotemark ~\cite{schmitt_sen12ms_2019} & 180,748 corresponding image triplets containing Sentinel-1 dual-pol SAR data, Sentinel-2 multi-spectral imagery, and MODIS-derived land cover maps, covering all inhabited continents during
all meteorological seasons. & image classification, \newline semantic segmentation, \newline data fusion & \cite{schmitt2018sen1} \\

MSAW\footnotemark ~\cite{shermeyer2020spacenet} & quad-pol X-band SAR imagery from Capella Space with 0.5 m spatial resolution, which covers 120 $\text{km}^2$ in the area of Rotterdam, the Netherlands. A total number of 48,000 unique building footprints are labeled with associated height information curated from the 3D Basis registratie Adressen en Gebouwen (3DBAG) dataset. & semantic segmentation &   \\

PolSF, Data\footnotemark, \newline Label\footnotemark ~\cite{xu2019polsf}  & The dataset includes PolSAR images of San Francisco from five different sensors. Each image was densely labeled to five or six classes, such as mountain, water, high-density urban, low-density urban, vegetation, developed, and bare soil. & image classification, \newline semantic segmentation  \newline data fusion &  \cite{cao2019pixel}\\

MSTAR\footnotemark ~\cite{ross1998standard} & 17,658 X-band very high resolution SAR images chips (patches) of 10 classes of different vehicles plus one class of simple geometric shaped target. SAR images of pure clutter are also included in the dataset. & object detection, \newline scene classification & \cite{chen2016target}  \cite{ding2016convolutional} \cite{gao2018deep}   \\

OpenSARShip 2.0\footnotemark ~\cite{li_opensarship_2017}& 34,528 Sentinel-1 SAR image chips of ships with the ship geometric information, the ship type, and the corresponding automatic identification system (AIS) information. & object detection, \newline scene classification  &  \cite{huang_opensarship_2018} \\

SAR-Ship-Dataset\footnotemark ~\cite{wang_sar_2019} & 43,819 Gaofen-3 or Sentinel-1 image chips of different ships. Each image chip has a dimension of 256 by 256 pixels in range and azimuth. & object detection, scene classification & \\


SARptical\footnotemark ~\cite{wang_fusing_2017} & 10,108 coregistered pairs of TerraSAR-X very high resolution spotlight image patch and UltraCAM aerial RGB image patch in Berlin, Germany. The coregistration is defined by the matching of the 3D position of the center of the image pair. & image matching & \cite{hughes2018identifying,wang2018sarptical}  \\

SEN1-2\footnotemark ~\cite{schmitt2018sen1} & 282,384 pairs of corresponding Sentinel-1 single polarization intensity, and Sentinel-2 RGB image patches, collected  across the globe. The patches are of dimension 256 by 256 pixels. &  image matching \newline data fusion & \cite{schmitt_sen12ms_2019} \\

\bottomrule[1.0pt]
\end{tabular}
\end{table*}

\subsection{\textbf{Image/Scene Classification}}

\begin{itemize}

    \item \textbf{So2Sat LCZ42} \cite{zhu2019so2sat}: So2Sat LCZ42  follows the local climate zones (LCZs) classification scheme. The dataset comprises 400,673 pairs of  dual-pol Sentinel-1 and multi-spectral Sentinel-2 image patches from 42 urban agglomerations, plus 10 additional smaller areas, across five continents. The image patches are hand-labelled into one of the 17 LCZ classes \cite{stewart_local_2012}. The Sentinel-1 image patches in this dataset contain both the geocoded single look complex image, as well as a despeckled Lee filtered variant. In particular, it is the first Earth observation dataset that provides a quantitative measure of the label uncertainty, achieved by letting a group of domain experts cast 10 independent votes on 19 cities in the dataset. The dataset therefore can be considered a large-scale data fusion and classification benchmark dataset for cutting-edge machine learning methodological developments, such as automatic topology learning, data fusion, and quantification of uncertainties.

    \item \textbf{OpenSARUrban} \cite{zhao_opensarurban_2020}: OpenSARUrban consists of 33,358  patches of Sentinel-1 dual-pol images covering 21 major cities in China. The dataset was \YW{manually} annotated according to a hierarchical classification scheme, with 10 classes of urban scenes at its finest level. Each image patch has a dimension of 100 by 100 pixels with a pixel spacing of 10 m (Sentinel-1 GRD product). This dataset can support deep learning studies of urban target characterization, and content-based SAR image queries.
    Fig. \ref{fig_opensarurban} shows some samples from the OpenSARUrban dataset.
\end{itemize}

\subsection{\textbf{Semantic Segmentation/Classification}}
\begin{itemize}
\item \textbf{SEN12MS} \cite{schmitt_sen12ms_2019}: SEN12MS was created based on its previous version SEN1-2 \cite{schmitt2018sen1}. SEN12MS consists of 180,662 triplets of dual-pol Sentinel-1 image patches, multi-spectral Sentinel-2 image patches, and MODIS land cover maps. The patches are georeferenced with a ground sampling distance of 10 m. Each image patch has a dimension of 256 by 256 pixels. We expect this dataset to support the community in developing sophisticated deep learning-based approaches for common tasks such as scene classification or semantic segmentation for land cover mapping. 

\item \textbf{MSAW} \cite{shermeyer2020spacenet}: The multi-sensor all-weather mapping (MSAW) dataset includes  high-resolution SAR data, which covers 120 $\text{km}^2$ in the area of Rotterdam, the Netherlands. The quad-polarized X-band SAR imagery from Capella Space with 0.5 m spatial resolution was used for the SpaceNet 6 Challenge. A total of 48,000 unique building footprints have been labeled with additional building heights.

\item \textbf{PolSF} \cite{xu2019polsf}: This dataset consists of PolSAR images of San Francisco from eight different sensors, including AIRSAR, ALOS-1, ALOS-2, RADARSAT-2, SENTINEL-1A, SENTINEL-1B, GAOFEN-3, and RISAT (data compiled by E. Pottier of IETR). Five of the eight images were densely labeled to five or six land use land cover classes in \cite{xu2019polsf}. These densely annotated images correspond to roughly 3,000 training patches of 128 by 128 pixels. Although the data volume is relatively low for deep learning research, this dataset is the only annotated multi-sensory PolSAR dataset, to the best of our knowledge. Therefore, we suggest that the creator of this dataset increase the number of annotated images to enable greater potential use of this dataset.

\end{itemize}

\subsection{\textbf{Object Detection}}
\label{ssec:mstar}
\begin{itemize}
    \item \textbf{MSTAR} \cite{ross1998standard}: The Moving and Stationary Target Acquisition and Recognition (MSTAR) dataset is one of the earliest datasets for SAR target recognition. The dataset consists of 
    total 17,658 X-band SAR image chips (patches) of 10 classes of vehicle plus one class of simple geometric shaped target. The collected SAR image patches are $128$ by $128$ pixels with a resolution of one foot in range and azimuth. In addition, 100 SAR images of clutter were also provided in the dataset.

    

   In our opinion, the number of image patches in this dataset is relatively low for deep learning models, especially considering the number of classes. In addition, this dataset represents a rather ideal \YW{and unrealistic} scenario: vehicles in the dataset are centered in the patch, and the clutter is quite homogeneous without disturbing signals. However, considering the scarcity of such datasets, MSTAR is a valuable source for target recognition.

    \item \textbf{OpenSARShip 2.0} \cite{li_opensarship_2017}: This dataset was built based on its previous version, OpenSARShip \cite{huang_opensarship_2018}. It contains 34,528 Sentinel-1 SAR image patches of different ships with automatic identification system (AIS) information. For each SAR image patch, the creators manually extracted the ship length, width, and direction, as well as its type by verifying this data on the Marine Traffic website \cite{li_opensarship_2017}. Among all the patches, about one-third is extracted from Sentinel-1 GRD products, and the other two-thirds are from Sentinel-1 SLC products. OpenSARShip 2.0 is one of the handful of SAR datasets suitable for object detection.
    \label{sec:opensarship}

    \item \textbf{SAR-Ship-Dataset} \cite{wang_sar_2019}: This dataset was created using 102 Gaofen-3 and 108 Sentinel-1 images. It consists of 43,819 ship chips of 256 pixels in both range and azimuth. These ships mainly have distinct scales and backgrounds. Therefore, this dataset can be employed for developing  multi-scale object detection models. 
    
    \item \textbf{FUSAR-Ship} \cite{xiyuefusar}: This dataset was created using space-time matched-up datasets of Gaofen-3 SAR images and ship AIS messages. It consists of over 5000 ship chips with corresponding ship information extracted from AIS messages, which can be used to trace back to each unique ship of any particular chip.
    
    \item \YS{AIR-SARShip-1.0/2.0 \cite{xian2019air}: This dataset comprises 31 (300) SAR images from the Geofen-3 satellite, which includes 1m and 3m resolution imagery with different imaging modes, such as spotlight and stripmap. There are more than ten object categories including ships, tankers, fishing boats and others. The scene types in the dataset include ports, islands, reefs and sea surfaces of different levels.}
\end{itemize}

\footnotetext[1]{\url{https://doi.org/10.14459/2018mp1483140}}
\footnotetext[2]{\url{https://www.tensorflow.org/datasets/catalog/so2sat}}
\footnotetext[3]{\url{https://doi.org/10.21227/3sz0-dp26}}
\footnotetext[4]{\url{https://mediatum.ub.tum.de/1474000}}
\footnotetext[5]{\url{https://spacenet.ai/sn6-challenge/}}
\footnotetext[6]{\url{https://www.ietr.fr/polsarpro-bio/san-francisco/}}
\footnotetext[7]{\url{https://github.com/liuxuvip/PolSF}}
\footnotetext[8]{\url{https://www.sdms.afrl.af.mil/index.php?collection=mstar}}
\footnotetext[9]{\url{http://opensar.sjtu.edu.cn/Data/Search}}
\footnotetext[10]{\url{https://github.com/CAESAR-Radi/SAR-Ship-Dataset}}
\footnotetext[11]{\url{https://www.sipeo.bgu.tum.de/downloads/SARptical_data.zip}}
\footnotetext[12]{\url{https://mediatum.ub.tum.de/1436631}}

\subsection{\textbf{Registration/Matching}}
\begin{itemize}

    \item \textbf{SARptical} \cite{wang_fusing_2017, wang2018sarptical}: The SARptical dataset was designed for interpreting VHR spaceborne SAR images of dense urban areas. This dataset consists of 10,108 pairs of corresponding very high resolution SAR and optical image patches, whose location is precisely coregistered in 3D. They are extracted from TerraSAR-X VHR spotlight images with resolution better than 1 m and UltraCAM aerial optical images of 20 cm pixel spacing, respectively. Unlike low and medium resolution images, high resolution SAR and optical images in dense urban areas have very distinct geometries. Therefore, in the SARptical dataset, the center points of each image pair are matched in 3D space via sophisticated 3D reconstruction and matching algorithms. The UTM coordinates of the center pixel of each pair are also made available publicly in the dataset. 
    This dataset contributes to applications of multi-modal data classification, and SAR optical images co-registering. However, we believe more training samples are required for learning complicated SAR optical image to image mapping.

    \item \textbf{SEN1-2} \cite{schmitt2018sen1}: The SEN1-2 dataset consists of 282,384 pairs of corresponding Sentinel-1 single polarization intensity and Sentinel-2 RGB image patches, collected from across the globe and throughout all meteorological seasons. The patches are of dimension 256 by 256 pixels. Their distribution over the four seasons is roughly even. SEN1-2 is the first large open dataset of this kind. We believe it will support further developments in the field of deep learning for remote sensing as well as multi-sensor data fusion, such as SAR image colorization, and SAR-optical image matching. 

\end{itemize}

\subsection{\textbf{Other Datasets}}
\begin{itemize}

    \item \textbf{Sample PolSAR images from ESA}: \url{https://earth.esa.int/web/polsarpro/data-sources/sample-datasets}. For example, the Flevoland PolSAR Dataset. Several works make use of this dataset for agricultural land use land cover classification. The authors of \cite{yu2012unsupervised,hoekman2003new, yang2010weakly} have manually labeled the dataset according to different classification schemes.

    \item \textbf{SAR Image Land Cover Datasets} \cite{dumitru_sar_2018}: This dataset is not publicly available. Please contact the creator.
    \item \textbf{Airbus Ship Detection Challenge}: \url{https://www.kaggle.com/c/airbus-ship-detection}.

\end{itemize}



\section{Conclusion and Future Trends}
\label{sec:con}
This paper reviews the current state-of-the-art of an important and under-exploited research field --- deep learning in SAR. Relevant deep learning models are introduced, and their applications in six application fields --- terrain surface classification, object detection, parameter inversion, despeckling, InSAR, and SAR-optical data fusion --- are analyzed in depth. Exisiting benchmark datasets and their limitations are discussed. In summary, despite early successes, full exploitation of deep learning in SAR is mostly limited by 1) the lack of large and representative benchmark datasets and 2) the defect of tailored deep learning models that make full consideration of SAR signal characteristics.  \par 

Looking forward, the years ahead will be exciting. Next generation spaceborne SAR missions will simultaneously provide high resolution and global coverage, which will enable novel applications such as monitoring the dynamic Earth. To retrieve geo-parameters from these data, development of new analytics methods are warranted. Deep learning is among the most promising methods. To fully unlock its potential in SAR/InSAR applications in this big SAR data era, there are several promising future directions:

\begin{itemize}
\item \textbf{Large and Representative Benchmark Datasets}: As summarized in this article, there is only a handful of SAR benchmarks, in particular when excluding multi-modal ones. For instance, in SAR target detection, methods are mainly tested on a single benchmark data set --- the MSTAR dataset, where only several thousands of target samples in total (several hundreds for each class) are provided for training. With respect to InSAR, due to the lack of ground truth, datasets are extremely deficient or nearly nonexistent. Large and representative expert-annotated benchmark datasets are in high demand in the SAR community, and deserve more attention.  
\item \textbf{Unsupervised Deep Learning}: To bypass the deficiencies in annotated data in SAR, unsupervised deep learning is a promising direction. These algorithms derive insights directly from the data itself, and work as feature learning, representation learning, or clustering, which could be further used for data-driven analytics. Autoencoders and their extensions, such as variational autoencoders (VAEs) and deep embedded clustering algorithms, are popular choices. With respect to denoising, in despeckling, the high complexity of SAR images and lack of ground truth make it infeasible to produce appropriate benchmarks from real data. Noise2Noise \cite{lehtinen2018noise2noise} is an elegant example of unsupervised denoising where the authors learn denoised data without clean data. Despite the nice visual appearance of the results, preserving details is a must for SAR applications.
\item \textbf{Interferometric Data Processing}: Since deep learning methods are initially applied to perception tasks in computer vision, many methods resort to transforming SAR images, e.g., PolSAR images, into RGB-like images in advance or focus only on intensities. In other words, the most essential component of a SAR measurement --- the phase information --- is not appropriately considered. Although CV-CNNs are capable of learning phase information and show great potential in processing CV-SAR images, only a few such attempts have been made \cite{zhang2017complex}. Extending CNN to complex domain, while being able to preserve the precious phase information, would enable networks to directly learn features from raw data, and would open up a wide range of SAR/InSAR applications.
%
%
\item \textbf{Quantification of Uncertainties}: Generally speaking, geo-parameter estimates without uncertainty measures are considered invalid in remote sensing. Appropriately trained deep learning models can achieve highly accurate predictions. Yet, they fail in quantifying the uncertainty of these predictions. Here, giving a statement about the predictive uncertainty, while considering both aleatoric uncertainty and epistemic uncertainty, is of crucial importance. The Bayesian deep learning community has developed a model-agnostic and easy-to-implement methodology to estimate both data and model uncertainty within deep learning models \cite{kendall2017uncertainties}, which are awaiting exploration by the SAR community. 
\item \textbf{Large Scale Nonlinear Optimization Problems}: The development of inversion algorithms should keep up the pace of data growth. Fast solvers are demanded for many advanced parameter inversion models, which often involve non-convex, nonlinear, and complex-valued optimization problems, such as compressive-sensing-based tomographic inversion, or low rank complex tensor decomposition for InSAR time series data analysis. In some cases, the iterations of the optimization algorithms perform similar computations as layers in neural networks, that is, a linear step followed by a non-linear activation (see for example, the iteratively reweighted least-squares approach). And it is thus meaningful to replace the computationally expensive optimization algorithms with unrolled deep architectures that could be trained from simulated data \cite{chen2018theo}.  

\item \textbf{Cognitive Sensors}: Radars --– and SARs in particular --– are very complex and versatile imaging machines. A variety of modes (stripmap, spotlight, ScanSAR, TOPS, etc.), swath-widths, incidence angles and polarizations can be programmed in near real-time. Cognitive radars go a giant step further; they adapt their operational modes autonomously to the environment to be imaged by an intelligent interplay of transmit waveforms, adaptive signal processing on the receiver side and learning. Cognitive SARs are still in their conceptual and experimental phase and are often justified by the stunning capabilities of the echo-location system of bats. In his early pioneering article \cite{Haykin2006} Haykin defines three ingredients of a cognitive radar: “1) intelligent signal processing, which builds on learning through interactions of the radar with the surrounding environment; 2) feedback from the receiver to the transmitter, which is a facilitator of intelligence; and 3) preservation of the information content of radar returns, which is realized by the Bayesian approach to target detection through tracking.”
Such a SAR could, e.g., perform a low resolution, yet wide swath, surveillance of a coastal area and in a first step detect objects of interest, like ships, in real-time. Based on these detections the transmit waveform can be modified such as to zoom into the region of interest and allow for a close-up look of the object and possibly classify or even identify it. Reinforcement (online) learning is part of the concept as well as fast and reliable detectors or classifiers (trained offline), e.g. based on deep learning. All this is edge computing; the learning algorithms have to perform in real-time and with the limited compute resources onboard the satellite or airplane. 

\end{itemize}
Last but not least, technology advances in deep learning in remote sensing would only be possible if experts in remote sensing and machine learning work closely together. This is particularly true when it comes to SAR. Thus, we encourage more joint initiatives working collaboratively toward deep learning powered, explainable and reproducible big SAR data analytics.

\bibliographystyle{IEEEtran}
\bibliography{DL_SAR}

 
\begin{IEEEbiography}[{\includegraphics[width=1in,height=1.25in,clip,keepaspectratio]{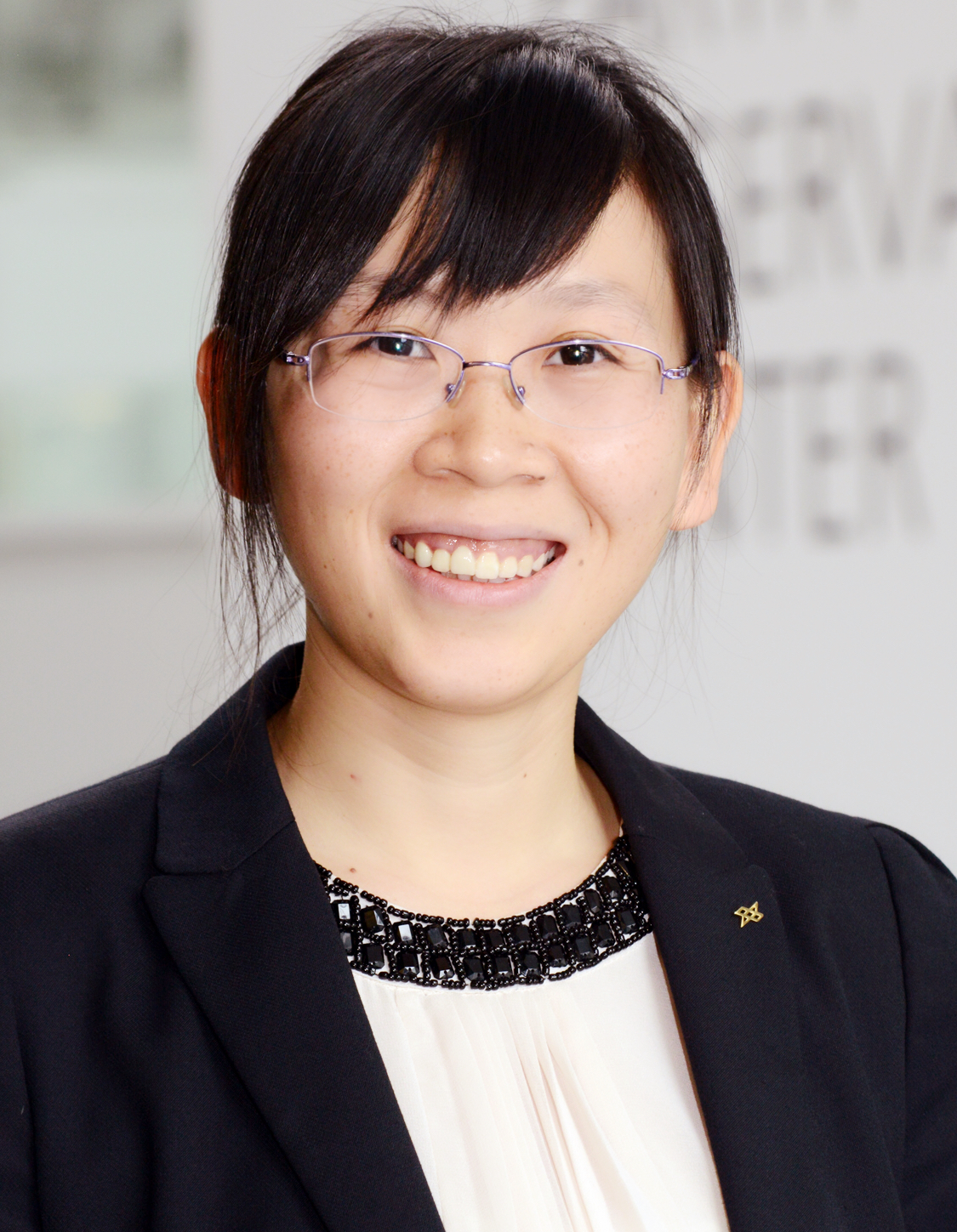}}]{Xiao Xiang Zhu}(S'10--M'12--SM'14--F'21) received the Master (M.Sc.) degree, her doctor of engineering (Dr.-Ing.) degree and her “Habilitation” in the field of signal processing from Technical University of Munich (TUM), Munich, Germany, in 2008, 2011 and 2013, respectively.
\par
She is currently the Professor for Data Science in Earth Observation (former: Signal Processing in Earth Observation) at Technical University of Munich (TUM) and the Head of the Department ``EO Data Science'' at the Remote Sensing Technology Institute, German Aerospace Center (DLR). Since 2019, Zhu is a co-coordinator of the Munich Data Science Research School (www.mu-ds.de). Since 2019 She also heads the Helmholtz Artificial Intelligence -- Research Field ``Aeronautics, Space and Transport". Since May 2020, she is the director of the international future AI lab "AI4EO -- Artificial Intelligence for Earth Observation: Reasoning, Uncertainties, Ethics and Beyond", Munich, Germany. Since October 2020, she also serves in the board of directors of the Munich Data Science Institute (MDSI), TUM. Prof. Zhu was a guest scientist or visiting professor at the Italian National Research Council (CNR-IREA), Naples, Italy, Fudan University, Shanghai, China, the University  of Tokyo, Tokyo, Japan and University of California, Los Angeles, United States in 2009, 2014, 2015 and 2016, respectively. Her main research interests are remote sensing and Earth observation, signal processing, machine learning and data science, with a special application focus on global urban mapping.

Dr. Zhu is a member of young academy (Junge Akademie/Junges Kolleg) at the Berlin-Brandenburg Academy of Sciences and Humanities and the German National  Academy of Sciences Leopoldina and the Bavarian Academy of Sciences and Humanities. She is an associate Editor of IEEE Transactions on Geoscience and Remote Sensing.
\end{IEEEbiography}

\begin{IEEEbiography}[{\includegraphics[width=1in,height=1.25in,clip,keepaspectratio]{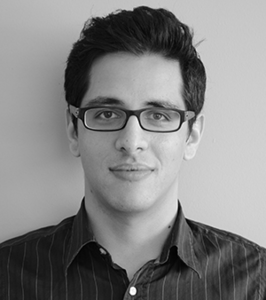}}]{Sina Montazeri} received the B.Sc. degree in geodetic engineering from the University of Isfahan, Isfahan, Iran, in 2011, the M.Sc. degree in geomatics from Delft University of Technology (TU Delft), Delft, The Netherlands in 2014, and the Ph.D. degree in radar remote sensing from the Technical University of Munich (TUM), Munich, Germany, in 2019 with a dissertation on Geodetic SAR Interferometry. In 2012, he spent two weeks with the Laboratoire des Sciences de l’Image, de l’Informatique et de la Télédétection, University of Strasbourg, Strasbourg, France, as a Junior Researcher working on thermal remote sensing. From 2013 to 2015, he was a Research Assistant with the Remote Sensing Technology Institute (IMF), German Aerospace Center (DLR), where he was involved in absolute localization of point clouds obtained from SAR tomography. From 2015 to 2019, he was a research associate with TUM-SiPEO and DLR-IMF working on automatic positioning of ground control points from multi-view radar images. He is currently a Senior Researcher with the department of EO Data Science of DLR-IMF focused on developing Machine Learning algorithms applied to radar imagery. His research interests include advanced InSAR techniques for deformation monitoring of urban infrastructure, image and signal processing relevant to radar imagery and applied machine learning.

Dr. Montazeri was the recipient of the DLR Science Award and the IEEE Geoscience and Remote Sensing Society Transactions Prize Paper Award, in 2016 and 2017, respectively for his work on Geodetic SAR Tomography.

\end{IEEEbiography}

\begin{IEEEbiography}[{\includegraphics[width=1in,height=1.25in,clip,keepaspectratio]{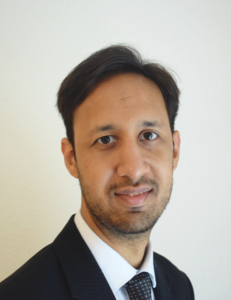}}]{Mohsin Ali}eceived his Bachelors in Computer Engineering degree from National University of Science and Technology(NUST), Islamabad, Pakistan in 2013 and Masters in Computer Science degree from University of Freiburg, Germany in 2018. Since April 2019 he is a PhD candidate at Earth Observation Center, DLR supervised by Prof. Dr. Xiaoxiang Zhu. His main research interest are uncertainty estimation in deep learning models for remote sensing applications.
\end{IEEEbiography}

\begin{IEEEbiography}[{\includegraphics[width=1in,height=1.25in,clip,keepaspectratio]{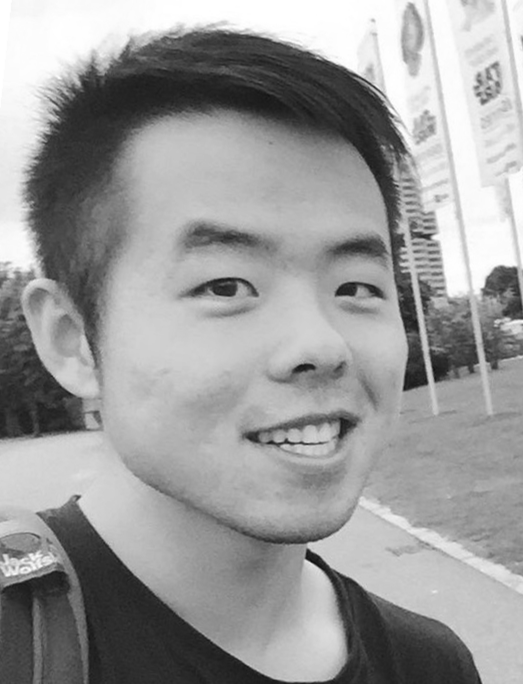}}]{Yuansheng Hua}
(S'18) received the Bachelor's degree in remote sensing science and technology from the Wuhan University, Wuhan, China, in 2014, and the Master's degree in Earth Oriented Space Science and Technology (ESPACE) from the Technical University of Munich (TUM), Munich, Germany, in 2018. 
\par
He is currently pursuing the Ph.D. degree with the German Aerospace Center (DLR), Wessling, Germany and the Technical University of Munich (TUM), Munich, Germany. 
\par
In 2019, he was a visiting researcher with the Wageningen University \& Research, Wageningen, Netherlands. His research interests include remote sensing, computer vision, and deep learning, especially their applications in remote sensing.
\end{IEEEbiography}

\begin{IEEEbiography}[{\includegraphics[width=1in,height=1.25in,clip,keepaspectratio]{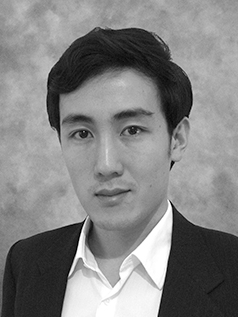}}]{Yuanyuan Wang} (S'08-M'11)
received the B.Eng. degree (Hons.) in electrical engineering from The Hong Kong Polytechnic University, Hong Kong, in 2008, and the M.Sc. and Dr. Ing. degree from the Technical University of Munich (TUM), Munich, Germany, in 2010 and 2015, respectively. In June and July 2014, he was a Guest Scientist with the Institute of Visual Computing, ETH Zürich, Zürich, Switzerland. He is currently with the Department of EO Data Science, Remote Sensing Technology Institute of the German Aerospace Center, Weßling, Germany, where he leads the working group \textit{Big SAR Data}. He is also a guest member of the Professorship of Data Science in Earth Observation, Technical University of Munich, Munich, Germany, where he supports the scientific management of ERC projects So2Sat (so2sat.eu) and AI4SmartCities (cordis.europa.eu/project/id/957467). His research interests include optimal and robust parameters estimation in multibaseline InSAR techniques, multisensor fusion algorithms of synthetic aperture radar (SAR) and optical data, nonlinear optimization with complex numbers, machine learning in SAR, and high-performance computing for big data. 

Dr. Wang serves as the reviewer for multiple IEEE GRSS and other remote sensing journals. He was one of the best reviewers of the IEEE Transactions on Geoscience and Remote Sensing in 2016. He is also an associate editor of the Geoscience Data Journal of the UK Royal Meteorological Society.
\end{IEEEbiography}
\begin{IEEEbiography}[{\includegraphics[width=1in,height=1.25in,clip,keepaspectratio]{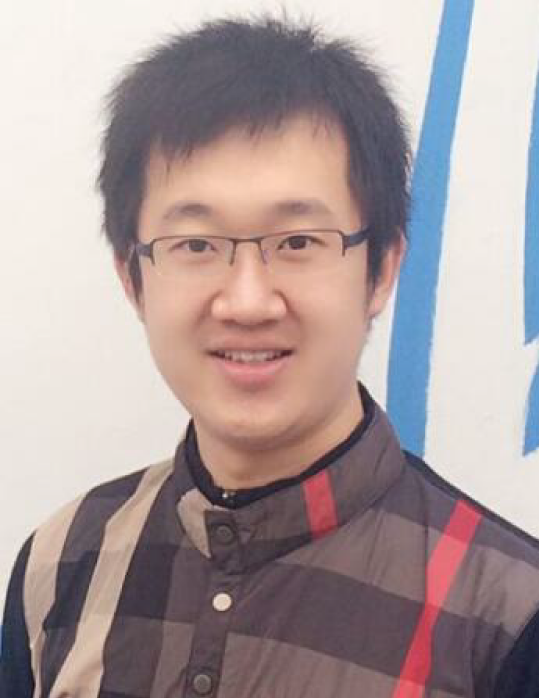}}]{Lichao Mou}
received the Bachelor's degree in automation from the Xi'an University of Posts and Telecommunications, Xi'an, China, in 2012, the Master's degree in signal and information processing from the University of Chinese Academy of Sciences (UCAS), China, in 2015, and the Dr.-Ing. degree from the Technical University of Munich (TUM), Munich, Germany, in 2020.
\par
He is currently a Guest Professor at the Munich AI Future Lab AI4EO, TUM and the Head of Visual Learning and Reasoning team at the Department ``EO Data Science'', Remote Sensing Technology Institute (IMF), German Aerospace Center (DLR), Wessling, Germany. Since 2019, he is an AI Consultant for the Helmholtz Artificial Intelligence Cooperation Unit (HAICU). In 2015 he spent six months at the Computer Vision Group at the University of Freiburg in Germany. In 2019 he was a Visiting Researcher with the Cambridge Image Analysis Group (CIA), University of Cambridge, UK. From 2019 to 2020, he was a Research Scientist at DLR-IMF.
\par
He was the recipient of the first place in the 2016 IEEE GRSS Data Fusion Contest and finalists for the Best Student Paper Award at the 2017 Joint Urban Remote Sensing Event and 2019 Joint Urban Remote Sensing Event.
\end{IEEEbiography}

\begin{IEEEbiography}[{\includegraphics[width=1in,height=1.5in,clip,keepaspectratio]{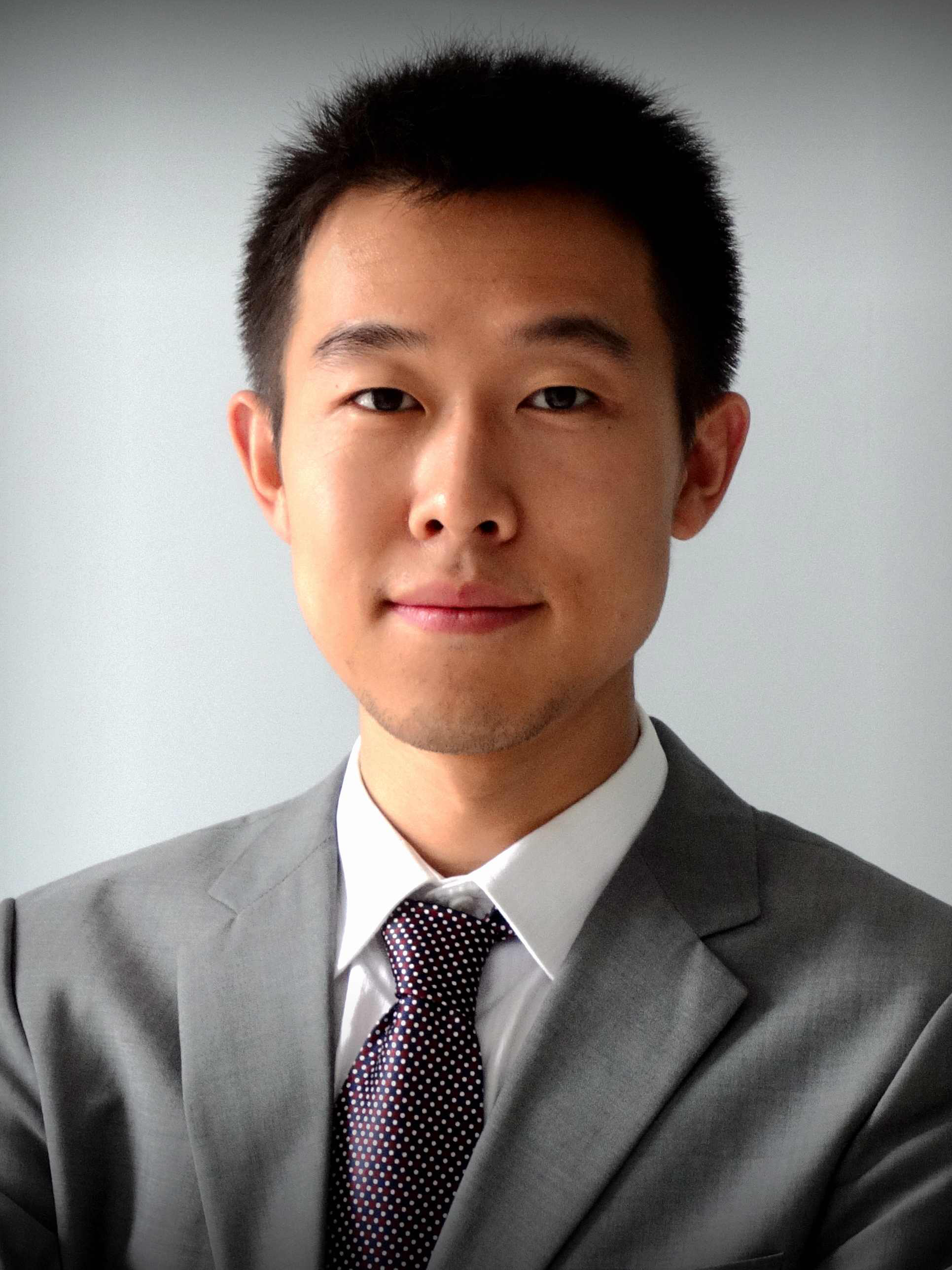}}]{Yilei Shi}
(M'18) received his Diploma (Dipl.-Ing.) degree in Mechanical Engineering, his Doctorate (Dr.-Ing.) degree in Engineering from Technical University of Munich (TUM), Germany. In April and May 2019, he was a guest scientist with the department of applied mathematics and theoretical physics, University of Cambridge, United Kingdom. He is currently a senior scientist with the Chair of Remote Sensing Technology, Technical University of Munich.

His research interests include computational intelligence, fast solver and parallel computing for large-scale problems, advanced methods on SAR and InSAR processing, machine learning and deep learning for variety data sources, such as SAR, optical images, medical images and so on; PDE related numerical modeling and computing.
\end{IEEEbiography}

\begin{IEEEbiography}[{\includegraphics[width=1in,height=1.25in,clip,keepaspectratio]{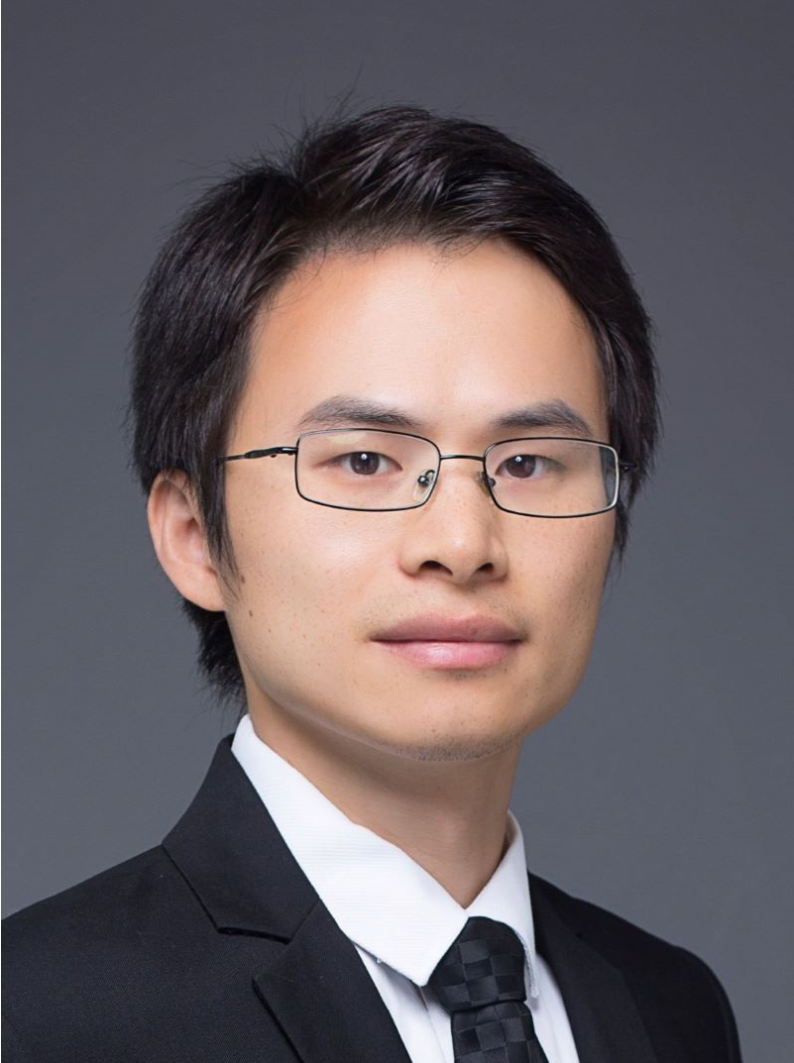}}]{Feng Xu}
(S'06-M'08-SM'14) received the B.E. (Hons.) degree in information engineering from Southeast University, Nanjing, China, in 2003, and the Ph.D. (Hons.) degree in electronic engineering from Fudan University, Shanghai, China, in 2008. From 2008 to 2010, he was a Post-Doctoral Fellow with the NOAA Center for Satellite Application and Research, Camp Springs, MD, USA. From 2010 to 2013, he was with Intelligent Automation Inc., Rockville, MD, USA, while he was partly with the NASA Goddard Space Flight Center, Greenbelt, MD, USA, as a Research Scientist. In 2012, he was selected into China's Global Experts Recruitment Program, and subsequently returned to Fudan University, Shanghai, China, in 2013, where he is currently a Professor with the School of Information Science and Technology and the Vice Director of the MoE Key Laboratory for Information Science of Electromagnetic Waves. He has authored more than 30 papers in peer-reviewed journals, co-authored two books, and holds two patents, among many conference papers. His research interests include electromagnetic scattering modeling, SAR information retrieval, and radar system development. 

Dr. Xu was a recipient of the second-class National Nature Science Award the IEEE Geoscience and Remote Sensing Society and the 2014 SUMMA Graduate Fellowship in the advanced electromagnetics area. He currently erves as the Associate Editor of the IEEE GEOSCIENCE AND REMOTE SENSING LETTERS. He is the Founding Chair of the IEEE GRSS Shanghai Chapter.
\end{IEEEbiography}

\begin{IEEEbiography}[{\includegraphics[width=1in,height=1.5in,clip,keepaspectratio]{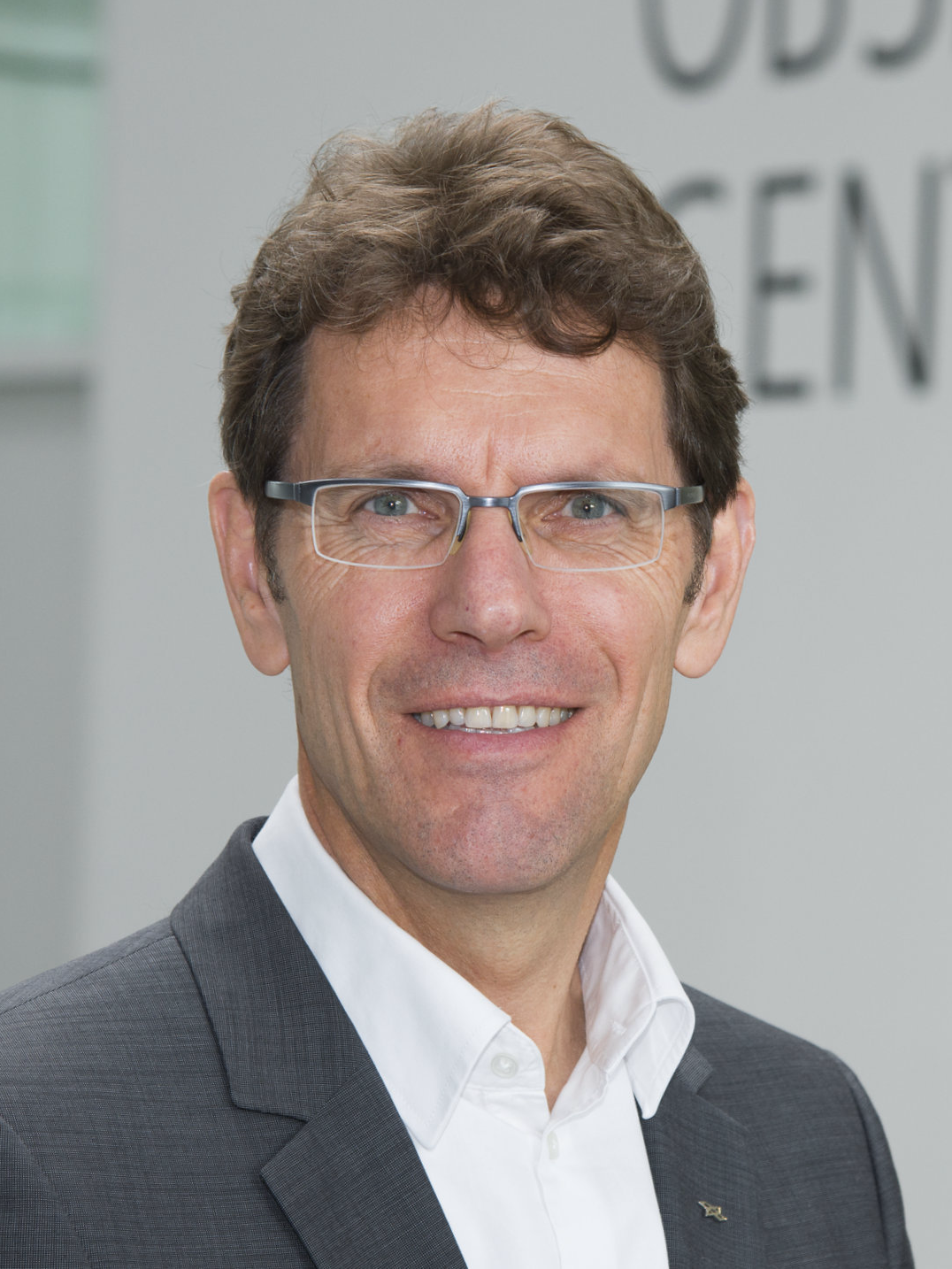}}]{Richard Bamler}
(M'95--SM'00--F'05) received his Diploma degree in Electrical Engineering, his Doctorate in Engineering, and his “Habilitation” in the field of signal and systems theory in 1980, 1986, and 1988, respectively, from the Technical University of Munich, Germany.

He worked at the university from 1981 to 1989 on optical signal processing, holography, wave propagation, and tomography. He joined the German Aerospace Center (DLR), Oberpfaffenhofen, in 1989, where he is currently the Director of the Remote Sensing Technology Institute.

In early 1994, Richard Bamler was a visiting scientist at Jet Propulsion Laboratory (JPL) in preparation of the SIC-C/X-SAR missions, and in 1996 he was guest professor at the University of Innsbruck. Since 2003 he has held a full professorship in remote sensing technology at the Technical University of Munich as a double appointment with his DLR position. His teaching activities include university lectures and courses on signal processing, estimation theory, and SAR. Since he joined DLR Richard Bamler, his team, and his institute have been working on SAR and optical remote sensing, image analysis and understanding, stereo reconstruction, computer vision, ocean color, passive and active atmospheric sounding, and laboratory spectrometry. They were and are responsible for the development of the operational processors for SIR-C/X-SAR, SRTM, TerraSAR-X, TanDEM-X, Tandem-L, ERS-2/GOME, ENVISAT/SCIAMACHY, MetOp/GOME-2, Sentinel-5P, Sentinel-4, DESIS, EnMAP, etc.

Richard Bamler’s research interests are in algorithms for optimum information extraction from remote sensing data with emphasis on SAR. This involves new estimation algorithms, like sparse reconstruction, compressive sensing and deep learning.
\end{IEEEbiography}

\end{document}